\documentclass[10pt,conference]{IEEEtran}
\usepackage{graphicx}
\usepackage[table,xcdraw]{xcolor}

\usepackage[T1]{fontenc}
\usepackage{amsmath,amssymb,amsfonts}
\usepackage{algorithmic}
\usepackage{mathtools}
\usepackage[linesnumbered,ruled,lined]{algorithm2e}
\usepackage{graphicx}
\usepackage[hidelinks]{hyperref}
\usepackage{academicons}
\usepackage{float}
\usepackage{textcomp}
\usepackage{subcaption}
\usepackage{mwe}
\usepackage{textcomp}
\usepackage{float}
\usepackage{wrapfig}
\usepackage{rotating}
\usepackage{tikz}
\usepackage{textcomp}
\usepackage{subcaption}
\usepackage{multicol}
\usepackage{graphicx}
\usepackage{dblfloatfix}
\usepackage{float}
\usepackage{amssymb}
\usepackage{gensymb}
\usepackage{array}
\usepackage{fancyhdr}
\usepackage{dirtytalk}
\usepackage{flexisym}
\usepackage{anyfontsize}
\DeclarePairedDelimiter{\ceil}{\lceil}{\rceil}
\usepackage{amsmath,amssymb,amsfonts}
\usepackage{algorithmic}
\usepackage{graphicx}
\usepackage{textcomp}
\usepackage{wrapfig}
\usepackage{scalerel}
\usepackage{tikz}
\newcommand{\ourmethod}{\textit{UniPreCIS}\;}

\begin{document}


\title{\ourmethod: A data pre-processing solution for collocated services on shared IoT}

\author{\IEEEauthorblockN{1\textsuperscript{st} Anirban~Das}
\IEEEauthorblockA{\textit{Indian Institute of Information Technology Guwahati, India} \\
anirban@iiitg.ac.in}\\
\and
\IEEEauthorblockN{2\textsuperscript{nd} Navlika~Singh}
\IEEEauthorblockA{\textit{Indian Institute of Technology Jodhpur, India} \\
singh.119@iitj.ac.in}\\
\and
\IEEEauthorblockN{3\textsuperscript{rd} Suchetana~Chakraborty}
\IEEEauthorblockA{\textit{Indian Institute of Technology Jodhpur, India} \\
suchetana@iitj.ac.in}\\
}

\maketitle

\title{\ourmethod: A data pre-processing solution for collocated services on shared IoT}





\begin{abstract}
Next-generation smart city applications, attributed by the power of Internet of Things (IoT) and Cyber-Physical Systems (CPS), significantly rely on the quality of sensing data. With an exponential increase in intelligent applications for urban development and enterprises offering sensing-as-a-service these days, it is imperative to provision for a shared sensing infrastructure for better utilization of resources. However a shared sensing infrastructure that leverages low-cost sensing devices for a cost effective solution, still remains an unexplored territory. 
A significant research effort is still needed to make edge based data shaping solutions, more reliable, feature-rich and cost-effective while addressing the associated challenges in sharing the sensing infrastructure among multiple collocated services with diverse Quality of Service (QoS) requirements. Towards this, we propose a novel edge based data pre-processing solution, named \ourmethod that accounts for the inherent characteristics of low-cost ambient sensors and the exhibited measurement dynamics with respect to application-specific QoS. \ourmethod aims to identify and select quality data sources by performing sensor ranking and selection followed by multimodal data pre-processing in order to meet heterogeneous application QoS and at the same time reducing the resource consumption footprint for the resource constrained network edge. As observed, the processing time and memory utilization has been reduced in the proposed approach while achieving upto 90\% accuracy which is arguably significant as compared to state-of-the-art techniques for sensing. The effectiveness of \ourmethod has been evaluated on a testbed for a specific use case of indoor occupancy estimation that proves its effectiveness. 
\end{abstract}




\section{Introduction}
\label{sec:introduction}
Rapid advent in Internet of Things (IoT) and Cyber Physical Systems (CPS) along with the emerging demand for ubiquitous intelligent services across all smart city verticals have strongly motivated the rise of \textit{ambient sensing}. For being less intrusive (unlike camera and wearables), ambient sensing~\cite{ranieri2021activity} has risen as a more practical solution in numerous use cases. Towards effective utilization of the resources and delivery of improved Quality of Services, next generation IoT is moving towards shared infrastructure for sensing~\cite{byabazaire2020using}. And in parallel to this, adoption of low-cost sensors~\cite{palmisani2021indoor, desouza2021distribution, placidi2022capacitive} is becoming more pervasive than ever across numerous domains. These two factors in turn significantly encourage the decoupling of IoT service providers from hardware deployment authority~\cite{theodorou2020network} as a cost effective, easy-to-manage and a quick-deployable solution for the service providers. The ideal implication of this sensing-as-a-service provisioning can be characterized mainly by two prominent features: \textit{heterogeneity} and \textit{redundancy}. Variety of intelligent services~\cite{preuveneers2019big} running at the edge depend on the quality of data generated from a massive amount of sensory sources of different types and deployed in abundance. This primarily aims to address two aspects: \textit{fault tolerance} and \textit{coverage}.
While the redundancy can offer resilience to error and heterogeneity has the power to uncover the feature-rich insights of the sensing context. But it is worth noting that redundancy add to the \textit{overhead} of data processing~\cite{LI2020111990}. 
For the sensors, the accuracy and reliability of measurements, which characterize the data quality, not only depend on the fabrication quality of the sensors or various environmental factors like dust, humidity, pressure, etc. but also the location of the deployment that feeds a running \textit{microservice}~\cite{PALLEWATTA2022121, dragoni2017microservices}. The level of noise associated with the sensory data could show a significant temporal variation due to inherent characteristics of MEMS (Micro-Electro Mechanical System) devices. Each service has certain QoS constraints such as timely inference, reliability, measurement accuracy etc. For instance, a smart Fire Alarm requires to maintain critical timeliness whereas a smart HVAC could be more aligned towards users' preferences and comfort. It is evident that these QoS constraints vary and their adequacy are overwhelmingly dependent on the underlying hardware infrastructure. 
However, with hardware deployment being separated, finding a set of perfectly matched hardware or ideally located sensors on the part of a service deployment authority, is not always feasible.

Traditionally, the redundancy, noise and error embedded in the sensory data is pre-processed~\cite{8816926} for effective utilization. However, considering the multiplicity, heterogeneity and resource scarcity towards the edge, the overall cost of streaming huge amount of multimodal data during continuous monitoring,~\cite{xu2022amnis} could be significant. 
Addressing the challenging task of data management towards the edge, enterprise level solutions such as Apache Kafka,
RabbitMQ,
Apache Spark,
Confluent,
Apache Storm,
etc. provide support for processing and managing heterogeneous IoT data in an edge-cloud architecture. These frameworks support various pre-processing functionalities like aggregation, compression, and fusion and are state of the art for real-life use cases on dedicated IoT infrastructure. However, 
to the best of our knowledge, no existing framework for edge-based data pre-processing offers the shaping of data in terms of quality and quantity towards the source for multiple services on \textit{shared sensory infrastructure}.

In this direction, we identify the following set of challenges:
\begin{itemize}

    \item[] {Challenge 1}: When redundant sources are available, how can the amount of data be reduced and poorly performing (less reliable) sources can be suppressed towards the extreme edge of the network?
    \item[] {Challenge 2}: When redundant sources are available, how can we choose the most appropriate set of sources for a specific service out of all the services relying on that particular modality?
    \item[] {Challenge 3}: While selecting a sensory source, only measurement accuracy may not be an adequate characteristic as a deciding factor, in such a case how to address multiple characteristics  such as response time, resolution, etc. of the sensory sources while selecting a set of sources?
    \item[] {Challenge 4}: While performing multi-factor based selection, it is evident that not all the characteristics of a sensory source are consistent over time, in such a case how to address this dynamicity?
    \item[] {Challenge 5}: When multiple services are relying on shared sensory sources, how can we take advantage of this while performing data processing to achieve certain benefits in terms of time and available resources?
    \item[] {Challenge 6}: How can the needs of modalities (fed by the sensory sources) of  individual depending services be formulated while preserving the necessary information, at the same time reducing the overall computation cost (in terms of resource utilisation)?

\end{itemize}

To address these questions, we introduce \ourmethod (Unified Pre -processing for Collocated IoT Services), a novel edge-based data pre-processing solution for shared IoT scenario. The key contribution of this work can be summarized as follows:
\begin{itemize}
    \item Towards data quality regulation, a novel multi-dimensional sensor ranking and selection approach based on dynamic sensor characteristics and application-specific QoS feedback has been introduced. The sensor ranking and selection addresses challenges 1-3 while modelling of dynamic characteristics address challenge 4.
    \item An edge based sensory data pre-processing framework has been proposed to fuse multimodal data and feed a set of collocated microservices corresponding to various smart applications sharing an IoT infrastructure. This aims to address challenge 5 and 6.
    \item The effectiveness is demonstrated by experimental results from testbed implementation. The framework is able to reduce the overall processing time by upto 23\% in our evaluation case. Whereas, the proposed sensor selection maintained the classification accuracy of a use case beyond 90\% without compromising the computation cost.
\end{itemize}
The rest of the paper is organized as follows. First, we provide the background information to motivate the problem. Then we introduce \ourmethod along with the system design and modelling. Next, we analyze the performance of \ourmethod with necessary experiments and results. After that we state the relevant literature and conclusion. And finally conclude with the relevant proofs provided in Appendix I.

\section{Background}
In a shared IoT infrastructure, 
a strong temporal correlation among different streams of generated data is the key requirement to leverage the power of multimodality and unfold the feature-rich deep insights of the sensing context. This section briefly introduces relevant properties of the sensory data, followed by a brief summary on different sensor characteristics and the external factors influencing them.

\subsection{Data and its Characteristics}
The sensors for continuous monitoring services usually generate data in the form of streams.
\subsubsection{Formalization of Data Streams}
The data generated from any \textit{particular} sensor device, $s$ can be expressed in the form of a set of tuples ($x_i$, $t_i$) where, $x_i$ is the $i^{th}$ measured value with $t_i$ as the timestamp. 
That said the generated stream of data, $X$ for the specific sensor $s$, can be expressed as, 
\begin{equation}
    X=\{(x_1, t_1),(x_2, t_{2}),...(x_k, t_{k})\}
    \label{eq:stream}
\end{equation}


\subsubsection{Fusion of Multiple Data Streams}
\label{kalmanfilter}



\begin{figure}[t]
\centering
\includegraphics[scale=.33, keepaspectratio]{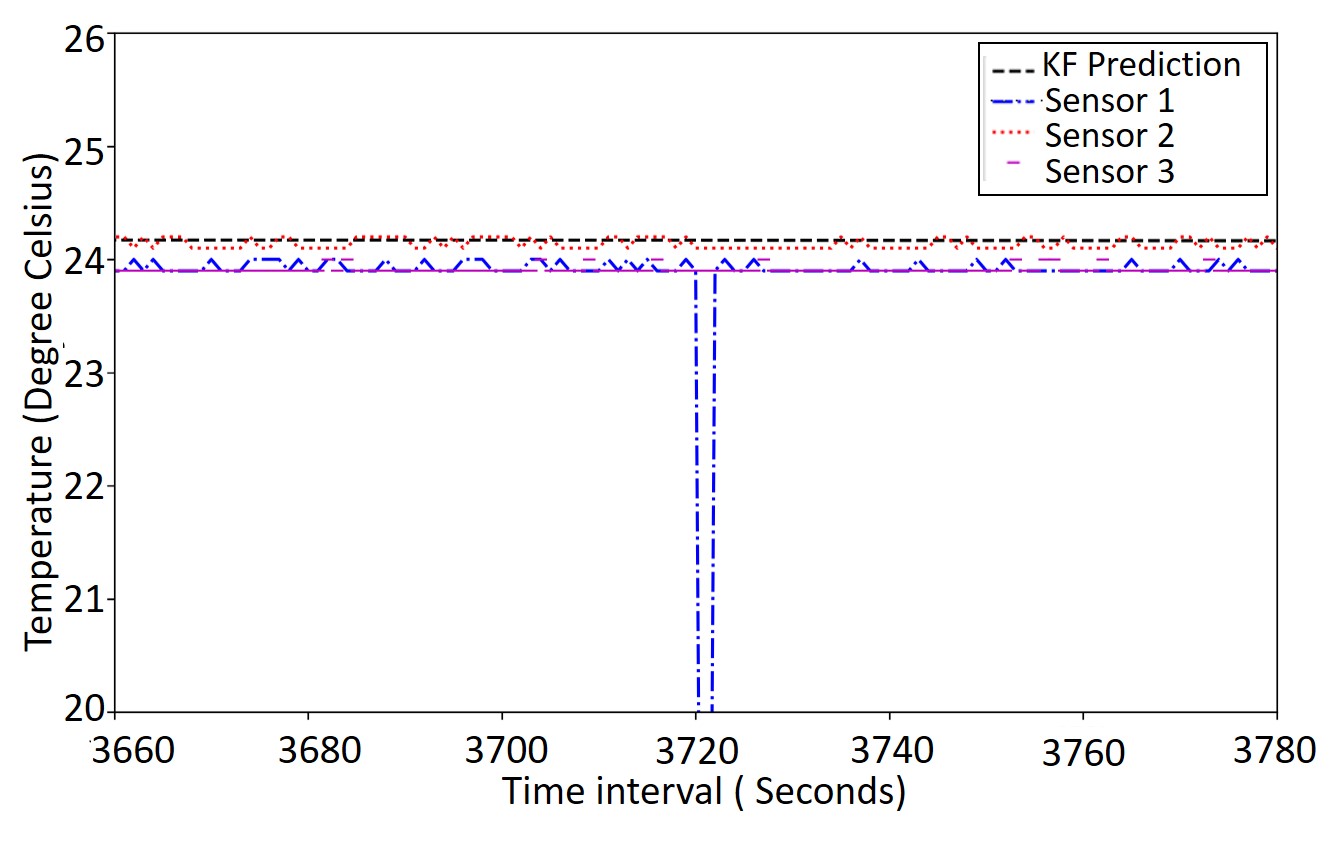}
\caption{Variation of temperature with time}
\label{fig:tempplot}
\end{figure}

A Linear \textit{Kalman Filter} (KF) can be used to fuse the data streams from an available set of sensors~\cite{8711204} to a single stream. 
Apart from fusing the streams, the KF can effectively filter out any noises generated by the sensors. For instance, Figure \ref{fig:tempplot} captures a scenario, where KF has been used to fuse streams from three temperature sensors monitoring an indoor lab space.

\begin{figure}[ht]
\centering
\includegraphics[scale=.33, keepaspectratio]{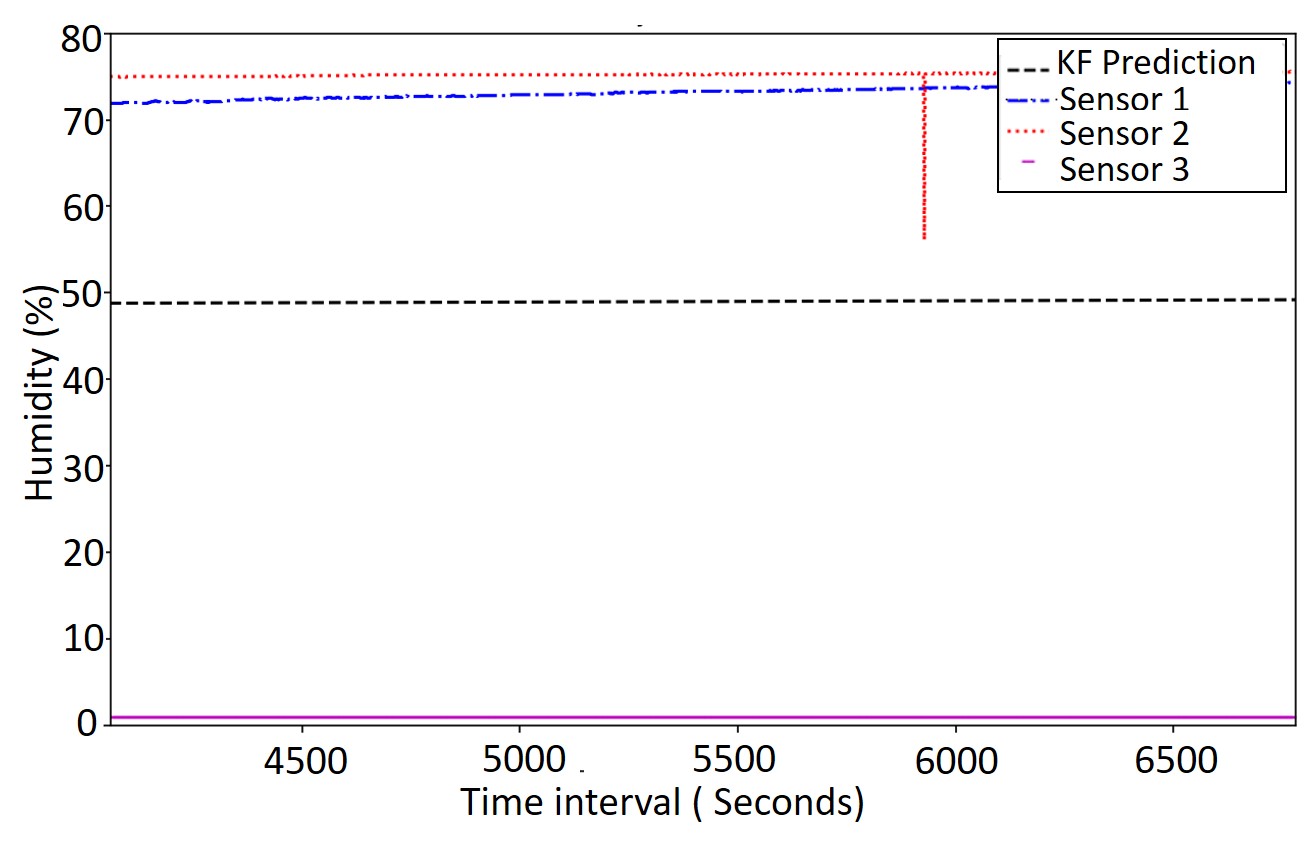}
\caption{Variation of humidity with time}
\label{fig:humidplot}
\end{figure}

KF in this case effectively minimizes the noise generated by sensor 1. It can be noted that since all the sensors were correctly functioning, the resultant curve from KF doesn't drift much from the individual sensor measurements. 
We now consider a scenario, as shown in Figure \ref{fig:humidplot} where three streams from three different humidity sensors has been plotted. In this case, \textit{sensor 3} was actually generating faulty readings. The effect of this incorrect readings is clearly visible as the output from the KF drifts far from the measurements of the rest of the two sensors.

\subsection{Sensor and its Characteristics}
\label{sensorcharacteristics}
\begin{figure}[t]
\centering
\includegraphics[scale=.38, keepaspectratio]{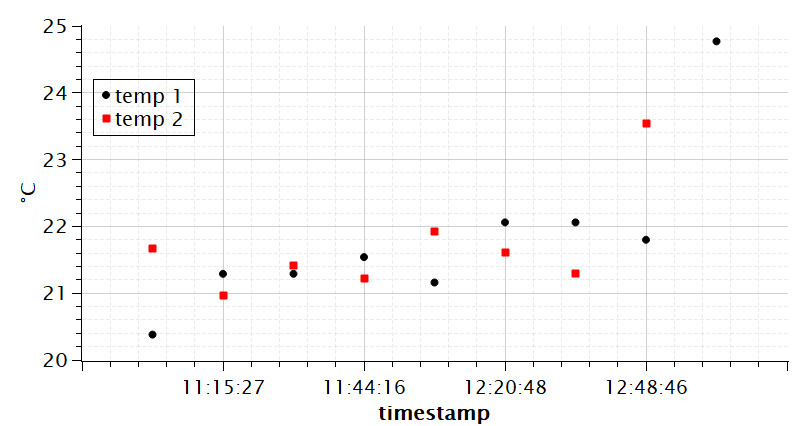}
\caption{Generated data points from two identical temperature sensors.}
\label{fig:data_plot}
\end{figure}

The low-cost sensors significantly boost affordability which is a major reason for the  rapidly growing research landscape of these sensors and their wide use in various applications \cite{shen2021temporal, giordano2021low, concas2021low}. However, there are certain characteristics where all are not equal. For instance, two low-cost temperature sensors (say DHT11 and DS18B20) can have different accuracy (+/- 2\degree C at 0 to 50\degree C and +/-0.5\degree C at -10 to 85\degree C respectively) and different measurement range (0 to 50\degree C and -55 to 125\degree C respectively)~\cite{petrich2020note}
due to the difference in make and model. Even for identical sensors, there can be missing values given a time interval. For instance, in Figure \ref{fig:data_plot}, the raw data~\cite{de2016benchmark} from two identical temperature sensors can be seen to have unequal quantity of data points generated given a time interval 10:00 to 13:00.

The sensor characteristics such as, accuracy, response time, range of measurements, residual energy, resolution, and so on play vital role to satisfy the QoS concerns of running services. For instance, a service may prefer a highly sensitive sensor to detect a minute change, whereas a different one may prefer a sensor with less sensitivity in order to avoid rapid fluctuations. These sensor characteristics may get affected by various external factors such as, interference, dust, human intervention, pressure, deployment etc. For instance, a PIR CO\textsubscript{2} sensor may respond inadequately if polled before its warm-up period. Apart from that there are many types of possible faults~\cite{jan2017sensor} that may affect the sensor's characteristics.
Certain scenarios may as well employ multiple instances of homogeneous sensors for improved resolution, enrichment of data or even mere extended coverage. 
But, computation on data from all the sensory sources every time is computationally expensive and can also be affected by incorrect measurements of a sensor as described in Figure~\ref{fig:humidplot}.


\section{System Model and Assumptions}
\begin{figure}[t]
\centering
\includegraphics[scale=.40, keepaspectratio]{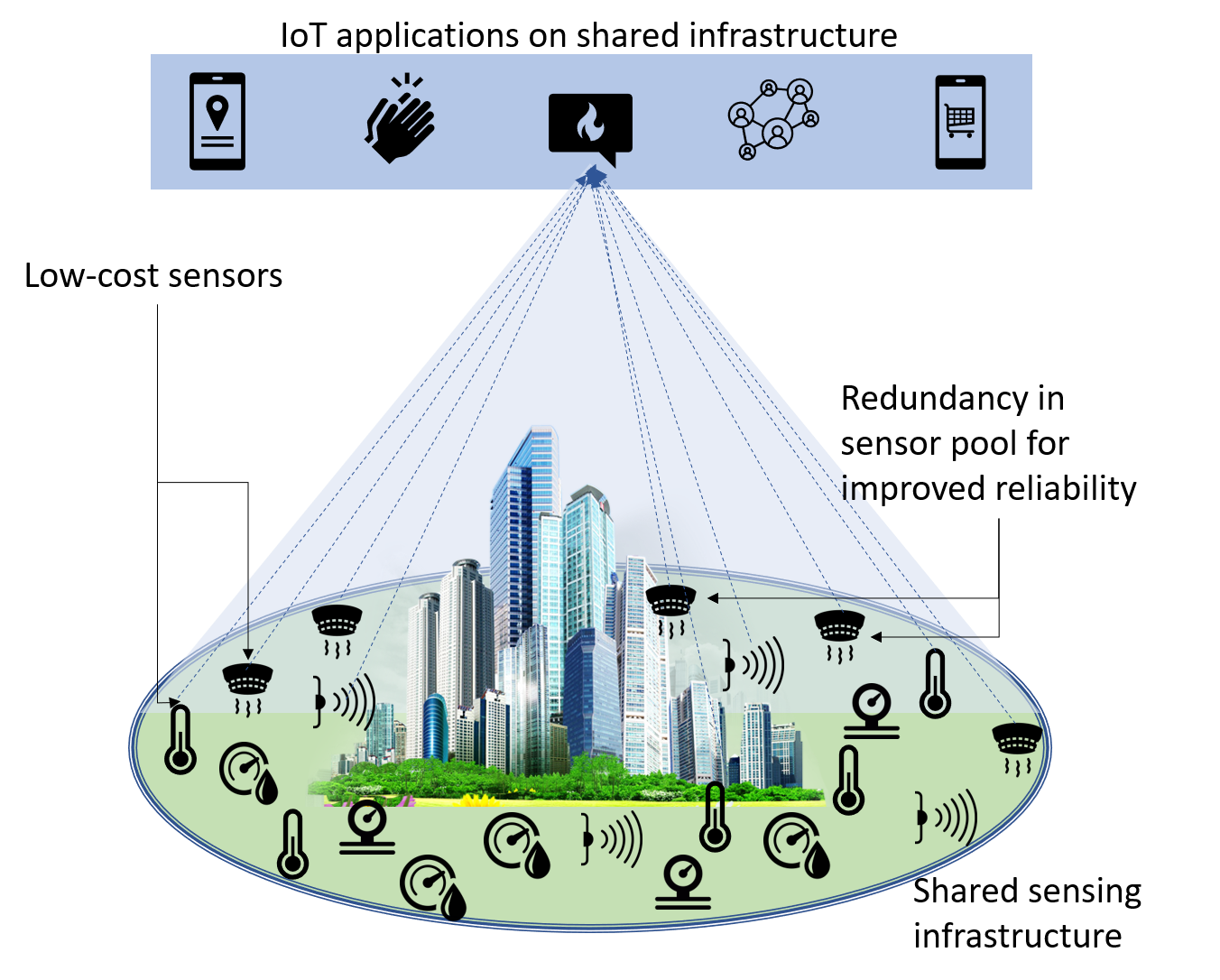}
\caption{An overview of the system model}
\label{fig:arch}
\end{figure}

A deployed set of sensors can be mapped dynamically to the stake-holder services while addressing the QoS awareness of these services. The system model for our solution can be represented by Figure~\ref{fig:arch}. A set of strategically deployed sensory sources measure the environment. The measurements are leveraged by the depending services while performing their respective computation. Towards this, we consider the following assumptions.
\begin{itemize}
    \item \textit{N} no. of heterogeneous services run on the edge. The services require data from heterogeneous sensors. 
    Multi-modal sensory data attributes are used to formulate the feature set meant to serve the stakeholder services, which is pretty generic. 
    \item A set of heterogeneous sensors periodically generate data to feed the available services. The sampling period of these sensors are independent. Time is the key entity stating the temporal correlation among the individual \textit{data points}.
    \item The characteristics of the available sensors like accuracy, response time, etc. can be defined by numerical metrics of which some are known a priori. 
\end{itemize}
The proposed methodology eventually optimizes the utilization of these sensory sources for the stakeholder services.

\begin{figure*}[ht]
\centering
\includegraphics[scale=.52, keepaspectratio]{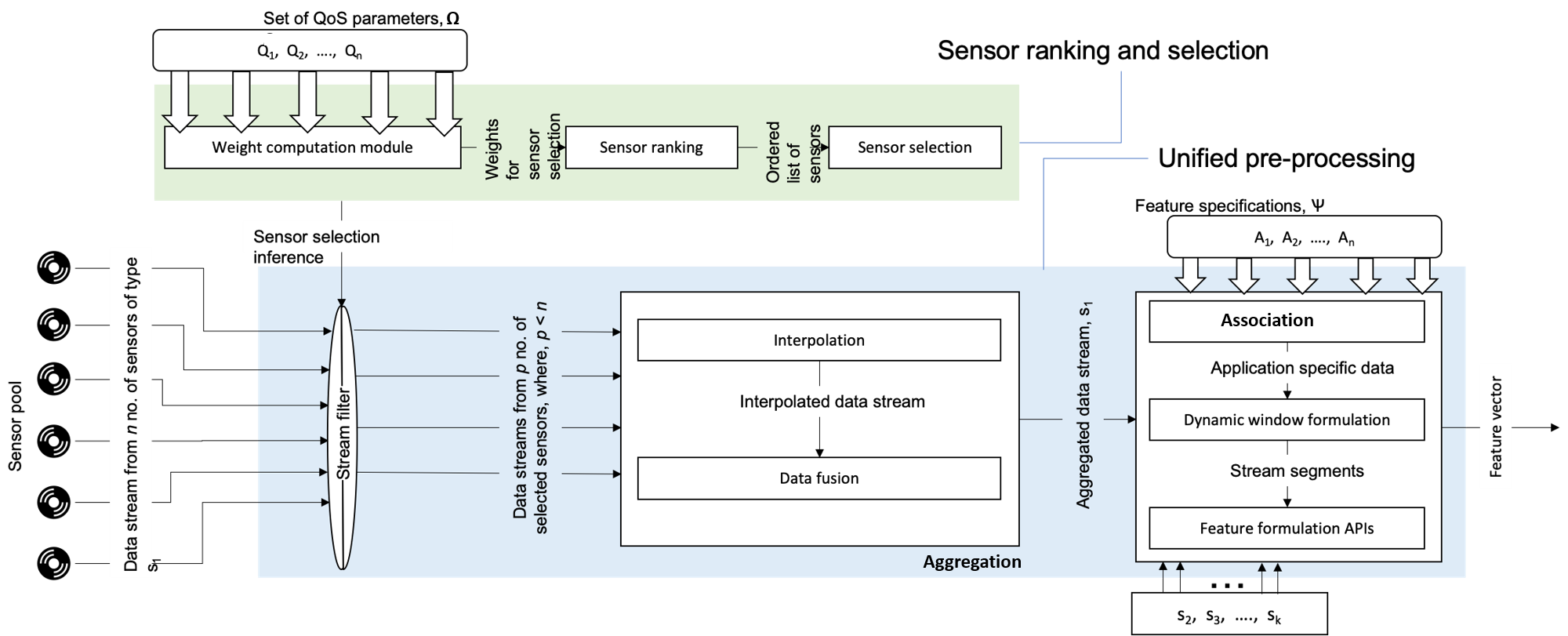}
\caption{Functional components of \ourmethod}
\label{fig:des}
\end{figure*}

\section{Proposed methodology}
In this section, we introduce \ourmethod a framework that eventually decouples the underlying sensor deployment from the relying services. The various functional components of \ourmethod is shown in Figure \ref{fig:des}. \ourmethod first isolates a set of optimal sensory sources to address the QoS awareness of the services. It then performs the pre-processing of the sensory measurements to generate service specific features in a unified manner. We detail these two key components below.

\subsection{Adaptive Sensor Ranking and Selection}
\label{rankingandsel}
With reference to the dynamic nature of the sensor characteristic as mentioned in Section~\ref{sensorcharacteristics}, let the $i^{th}$ sensor, of type $u$ can be represented as,
\begin{equation}
    u_i^t = [\alpha_i^t, \beta_i^t, \gamma_i^t, \epsilon_i^t, \kappa_i^t]
    \label{uvector}
\end{equation}
where, $u_i^t$ is the vector representation at time instance, $t$. In this case, we limit the no. of sensor properties to five for simplicity and refer them as, $\alpha$ for accuracy, $\beta$ for reliability, $\gamma$ for resolution, $\epsilon$ for response time and  $\kappa$ for range of measurements. We assume that $\alpha$ and $\beta$ are dynamic properties which vary over time. Whereas, the rest adhere to the sensor documentations.
We propose modelling of the two sensor attributes in an effective and light-weight manner.
\subsubsection{Modelling of Accuracy}
We combine a low-cost majority voting based model with \textit{dynamic inference} for assigning weights to a set of collocated homogeneous sensors.

Let a set of $n$ collocated sensors, $S=\{ s_1, s_2, s_3,…,s_n \}$ sense the same environment. 
Let k, ($1 \leq k \leq n$) be the no. of sensors generating near-accurate value.
Let $x_1, x_2, x_3,…,x_n$ be the measurements performed by the sensors within the time interval $\delta_t$ respectively.
Therefore, given this set, \\$X=\{ x_1, x_2, x_3,…,x_n \}$, the task is to find the sensors with the near optimal measurements. 
Assumptions:
\begin{itemize}
    \item $k=\ceil{\frac{n}{2}}$ (threshold condition) is the minimum no. of sensors who are generating optimal values.
	\item It’s a single dimensional linear system as the sensors are homogeneous.
	\item The number of sensors are limited to a few hundreds.
	\end{itemize}
The sensors generating near optimal values will report measurements close to each other. Therefore, there exists a fuzzy correlation between each pair of the measurements. We first normalise \cite{ping2007voting} the values as,
\[x_i = \frac{x_i - \overline{x}}{\delta}\]
\[\text{where, \:\:\:\:\:\:\:}  \overline{x} = \frac{1}{n} \sum_{i=1}^{n} x_i  \text{ \:\:and \:\:}    \delta= \sqrt{\frac{1}{n-1}\sum_{i=1}^{n}{(x_i-\overline{x})}^2}\] 
Now, we define fuzzy degree of membership function $f(.)$ as,
\[f(.)=e^{-\frac{{(x_i-x_j)}^p}{c}}\]
where, $p$ ($p\in \mathbb{N}$, $p>0$, $p\%2=0$) and $c$ ($c\in \mathbb{N}$, $c>0$) are two constants for scaling. Now we form a matrix, $M$ with the degree of membership between each pair of measurements as,
\[M=\begin{bmatrix}
e^{-\frac{{(x_1-x_1)}^p}{c}} & e^{-\frac{{(x_1-x_2)}^p}{c}} & ... &e^{-\frac{{(x_1-x_n)}^p}{c}}\\
e^{-\frac{{(x_2-x_1)}^p}{c}} & e^{-\frac{{(x_2-x_2)}^p}{c}} & ... &e^{-\frac{{(x_2-x_n)}^p}{c}}\\
\vdots & \ddots & \\
e^{-\frac{{(x_n-x_1)}^p}{c}} & e^{-\frac{{(x_n-x_2)}^p}{c}} & ... &e^{-\frac{{(x_n-x_n)}^p}{c}}\\
\end{bmatrix}\]

Let the matrix $M$ be of the form:

\[M=\begin{bmatrix}
U(x_1,x_1) & U(x_1,x_2) & ... &U(x_1,x_n)\\
U(x_2,x_1) & U(x_2,x_2) & ... &U(x_2,x_n)\\
\vdots & \ddots & \\
U(x_n,x_1) & U(x_n,x_2) & ... &U(x_n,x_n)\\
\end{bmatrix}\]

Of course, $M$ can be optimally represented by an upper or lower triangular matrix due to repetition of the members.
Now, for each $x_i$, we compute $C_{x_i}$~\footnote{Detailed in APPENDIX I} as,
\begin{equation}
C_{x_i}=\sum_{j=1}^{n}U(x_i,x_j)
\label{calcc}
\end{equation}
After that we sort all the $C_{x_i}$s in descending order. Finally, we isolate \textit{first} $k$ elements from the sorted list. Let the sorted list is renamed (each terms) as,
\[
 \{C_{m}, C_{m+1},..., C_{m+k-1}\}
\]
where, $C_{m}$ is the first element of the sorted list. Since, the values $C$ are computed on $x_m$s, each value of $C_m$ has a corresponding $x_m$ on which it was computed (as in eq. \ref{calcc}). To find the optimal set of $x_m$s, we now take the corresponding $x_m$s of the sorted list of $C$s and formulate, $\Bar{X}$

\begin{equation}
\Bar{X}=\{x_m, x_{m+1},..., x_{m+k-1}\}
\label{xbar}
\end{equation}

Here, $\Bar{X} \subseteq X$ and the elements are sorted in descending order. Once the optimal set of sensors is formulated, we calculate the true state of the sensors by invoking a linear KF on $\Bar{X}$. Let $y$ be the measurement calculated by KF($\Bar{X}$). In that case, the value for accuracy of the sensor, $s_i$, $\alpha_i$ is given by,
\[\alpha_i=e^{-|x_i-y|}\]
The ideal value in this case is 1 and represents the highest accuracy. The overall idea for the assignment of accuracy can be presented by Algorithm \ref{algo:accuracy}.

\begin{algorithm}[ht]
\caption[Dynamic assignment of accuracy]{\small {Given the set of measurements $X$=\{$x_1$, $x_2$,...$x_n$\} from collocated homogeneous sensors $S$=\{$s_1$, $s_2$,...$s_n$\}} respectively}
\label{algo:accuracy}
\scriptsize
\algsetup{indent=2em,linenodelimiter=.}
\begin{algorithmic} 
\renewcommand{\algorithmiccomment}[1]{/*#1*/}
\STATE $\Bar{x} \gets mean(X)$\\
\STATE $\delta \gets SD(X)$\\
\STATE $x\textsubscript{i} \gets \frac{x\textsubscript{i}-\Bar{x}}{\delta}$ \hspace{20pt}\text{\COMMENT{standardize $x_i$}}\\
\For{$ \forall i \in X $}
{
\For{$ \forall j \in X $}
{  
    $t \gets (x_i-x_j)\textsuperscript{ p}$\\
    $U(x_i,x_j) \gets \exp{(-\frac{t}{c})}$ \hspace{5pt}\text{\COMMENT{Degree of membership: f($x_i$)}}\\
}
}
\STATE $\text{Formulate }M$
\STATE $C_{x_i} \gets \sum_{j=1}^n U(x_i, x_j)$ \hspace{20pt}\text{\COMMENT{Calculate cumulative score}}
\STATE{formulate $\Bar{X}$}
\STATE $y \gets KF(\Bar{X}) $ \hspace{20pt}\text{\COMMENT{calculate Kalman Filter estimated true state}}\\
\STATE $\text{accuracy of ith sensor,}(s_i) \gets e^{-|x_i-y|}$\\
\end{algorithmic}
\end{algorithm}

\subsubsection{QoS Oriented Modelling of Reliability}

In a mission critical scenario, a highly reliable sensor may be preferred over a highly accurate sensor. Therefore, if the system only depends on accuracy, at times it may end up collapsing while handling an event. Although, reliability can be modeled in various ways, here we ascertain reliability at the data level (Section~\ref{sec:reliability}). Thus, the assignment of weight corresponding to reliability, to a sensor is performed based on the consistency in generating optimal values by the sensor over a certain period of time. To achieve this, we take the advantage of multiple sensors. The measurements included in the set $\Bar{X}$ in the expression \ref{xbar}  can be referred as near optimal values. That said we have the following assumptions:
\begin{itemize}
    \item For all, $x_i\in \Bar{X}$, $x_i$ is a value which is close to \textit{correct}. And for all $x_i \notin \Bar{X}$, $x_i$ is a reported value which is likely to be an incorrect measurement.
    \item The incorrect value measured is irrespective of hardware issues and noise.
    \item Cumulative QoS demands specify the expected data granularity within the time interval, $t_1$ and $t_2$ 
\end{itemize}
A sensor may not report a QoS demanded granularity at a certain interval of time, $t+\delta_t$ due to inherent configuration thus skipping a value at a required instance.

Let $n$ be the no. of expected values within the time interval $]t_1,t_2]$

Let $q$ be the no. of \textit{correct values} generated within $]t_1,t_2]$

From this, we define, $\lambda$ within $]t_1,t_2]$

\begin{equation}
\lambda =
    \begin{cases}
      \frac{q}{n} & \text{when q \textless n} \\
      1 & \text{otherwise}\\
    \end{cases}       
\end{equation}

$\therefore$ probability of $s_i$ during this time interval to be perfectly functioning, $P^t$ can be expressed using a Poisson distribution \cite{kuo2003optimal} defined as,
\[P^t=\frac{(\lambda t)^n}{n!}e^{-\lambda t}\]
where, $t=t_2-t_1, t_2>t_1$
Again, considering, $\lambda$ computed on various time intervals, $]t_i, t_j]$, $t_j>t_i$ of variable lengths, the cumulative probability that the sensor, $s_i$ is correct can be computed within $]t_0, T]$ (where, $T>t_0$) as,
\[P^T=\frac{(\int_{t_0}^{T} \lambda(t) \,dt)^n}{n!}e^{-\int_{t_0}^{T} \lambda(t) \,dt}\]
Here, $t=T-t_0$ is the interval of measurements. Since, both $0\leq p^t \leq 1$ and $0\leq P^T \leq 1$, either of these can be used as the assigned reliability metric for the sensor $s_i$. That said, we have,
\[\beta_i=p^t\: or\: \beta_i=P^T\]
The assignment of weight corresponding to reliability can thus be expressed in terms of Algorithm \ref{algo:reliability}

\begin{algorithm}[ht]
\caption[Dynamic assignment of reliability]{\small {Given the set of optimal sensors $\Bar{S_i}=\{s_j, s_{j+1},..., s_R\}$ at time instance $i$ from collocated homogeneous sensors $S$=\{$s_1$, $s_2$,...$s_n$\} respectively}}
\label{algo:reliability}
\scriptsize
\algsetup{indent=2em,linenodelimiter=.}
\begin{algorithmic}[1] 
\renewcommand{\algorithmiccomment}[1]{/*#1*/}
\STATE INITIALIZE $p^t_i = 0$\hspace{10pt}\COMMENT{Reliability measure for sensor $s_i$}\\
\For{$ \forall i \in [t_1, t_2] $}
{
    \IF{$s_i \in \Bar{S_i}$}
        \STATE $q_{s_i} \gets q_{s_i} +1$\\
    \ENDIF
}
\STATE $\lambda_{s_i} \gets \frac{q_{s_i}}{n}$\\
\STATE $t \gets t_2 - t_1$\\
\STATE $p^t_i \gets \frac{{(\lambda_{s_i} t)}^n}{n!}e^{-\lambda t}$ 
\end{algorithmic}
\end{algorithm}

\subsubsection{Multi-dimensional sensor ranking}
For all other static attributes $\Theta_i$, the values can be assigned between [0, 1] appropriately. For simplicity, given the documented specification, $\theta_i$, we state $\Theta_i$ as,
\[\Theta_i=e^{-\theta_i}\]
Once, the value of $\alpha,  \beta, \gamma, \epsilon$ and $\kappa$ are assigned for each $s_i$, the next is to perform multi-dimensional ranking of the sensors. For that we consider an ideal sensor as the \textit{utopia vector} of reference. As stated in eq. \ref{uvector}, let $v$ be the utopia vector of reference and is constituted as,
\begin{equation}
    v^t = [\alpha^t, \beta^t, \gamma^t, \epsilon^t, \kappa_i^t]
    \label{vvector}
\end{equation}
$\alpha,  \beta, \gamma, \epsilon$ and $\kappa$ will have their maximum possible values as unity. However, the QoS specifications of the stakeholder services specify the characteristics to formulate the vector $v$.
When $t$ is common for both $u$ and $v$, we can rewrite $u$ for a specific sensor and $v$ as,
\[u = [a_1^u, a_2^u, a_3^u, a_4^u, a_5^u]\text{,\:\:\:\:\: }
v = [b_1^v, b_2^v, b_3^v, b_4^v, b_5^v]\]
Here, $a_i^u$ and $b_i^v$ are one of the attributes of $u$ and $v$ respectively. The selection of the most prominent sensor, expressed as the \textit{compromised vector}, based on all associated attribute specifications, can be performed by measuring the Euclidean distance between the positions of each sensor from the utopia sensor $v$ as,
\begin{equation}
    \text{Euclidean component, $d_M$(u, v)} = \sqrt{\sum\limits_{i=1}^{m}(a_i^u-b_i^v)^2}
    \label{euclinoweight}
\end{equation}
In the multi-dimensional vector space, a directional component can be incorporated as,
\begin{equation}
    \text{Cos component, $d_A$(u, v)}= 1 - \frac{\sum\limits_{i=1}^{m}|a_i^u|\times |b_i^v|}{\sqrt{\sum\limits_{i=1}^{m}(a_i^u)^2\times\sum\limits_{i=1}^{m}(b_i^v)^2}}
    \label{cosinenoweight}
\end{equation}
Here, $m$ is the number of dimensions. Again, if the Qos specifications of the stakeholder services demand priority of one characteristic over another, a weighted version of $d_M$ and $d_A$ can be expressed as,
\begin{equation}
    \text{ $d_M$(u, v)} = \sqrt{\sum\limits_{i=1}^{m}w_i(a_i^u-b_i^v)^2}
    \label{eucliweight}
\end{equation}

\begin{equation}
    \text{$d_A$(u, v)}= 1 - \frac{\sum\limits_{i=1}^{m}|w_ia_i^u|\times |w_ib_i^v|}{\sqrt{\sum\limits_{i=1}^{m}(w_ia_i^u)^2\times\sum\limits_{i=1}^{m}(w_ib_i^v)^2}}
    \label{cosineweight}
\end{equation}
where, $w_i$ is the weight assigned signifying the priority level of the particular attribute. Therefore, combining, Eq. (\ref{euclinoweight}) and Eq. (\ref{cosinenoweight}) or Eq. (\ref{eucliweight}) and Eq. (\ref{cosineweight}), the direction aware distance, $d_{MA}$ between each vector $u$ and the utopia vector $v$ can be expressed as~\cite{gu2017new},
\begin{equation}
    \text{$d_{MA}$(u, v)}= \sqrt{\xi_M^2(d_M(u,v))^2+\xi_A^2(d_A(u,v))^2}
\end{equation}
where, $\xi_M$ and $\xi_A$ are a pair of scaling coefficients, which can be expressed as the inverse of the average of $d_M$ and $d_A$ respectively for equal importance to both the components.
\[
    \xi_M=\frac{m}{\sum\limits_{i=1}^{m}d_{M_i}}
\text{,\:\:\:\:\: }
    \xi_A=\frac{m}{\sum\limits_{i=1}^{m}d_{A_i}}
\]

Finally, the ranking of the sensors $s_i$ can be performed by sorting the sensors as per the value of their distances as measured by $d_{MA}$. The ranking and selection can be expressed as, Algorithm \ref{algo:selection}. The selected streams from this can be further aggregated in the aggregation module.
\begin{algorithm}[ht]
\caption[Sensor selection]
{\small {Given the utopia sensor, $v$ as [$b_1^v, b_2^v, b_3^v, b_4^v, b_5^v$] and a set of homogeneous sensors $S$, each represented by $u$ as [$a_1^u, a_2^u, a_3^u, a_4^u, a_5^u$]}}
\label{algo:selection}
\scriptsize
\algsetup{indent=2em,linenodelimiter=.}
\begin{algorithmic}
\renewcommand{\algorithmiccomment}[1]{/*#1*/}
\STATE INITIALIZE $d_{MA}(u_i) = 0 $\\
\For{$ \forall i \in S $}
{   
    $d_M(u_i, v) \gets $\text{Euclidean distance} $(u_i, v)$\\
    $d_A(u_i, v) \gets $\text{Cosine distance} $(u_i, v)$ \\
    $d_{MA}(u_i) \gets $\text{sqrt}$(\xi_M^2(d_M)^2+\xi_A^2(d_A)^2)$\\
}
\STATE INITIALIZE selected sensor $SEL = null $\\
\For{$ \forall i \in S$ }
{
\For{$ \forall j \in S $}
{ 
    \IF{$d_{MA} (u_i) < d_{MA} (u_j)$}
        \STATE $SEL \gets d_{MA} (u_i)$
    \ELSIF{$d_{MA} (u_j) < d_{MA} (u_i)$}
        \STATE $SEL \gets d_{MA} (u_j)$
    \ENDIF
}
}
\end{algorithmic}
\end{algorithm}

\subsection{Unified Data Pre-processing}
\begin{figure}[t]
\centering
\includegraphics[scale=.45, keepaspectratio]{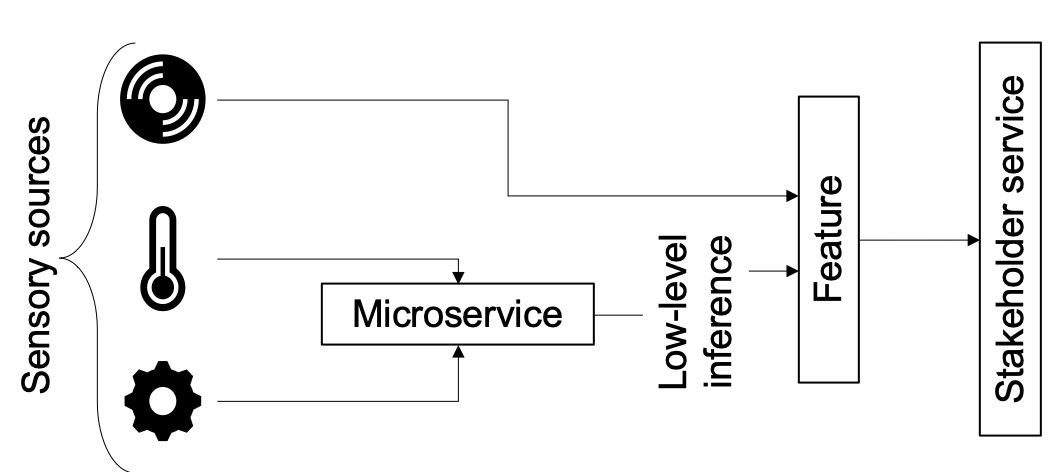}
\caption{Piping of low level inference for feature formulation.}
\label{fig:lowinfer}
\end{figure}

Unified pre-processing is performed to reduce redundant pre-processing steps in a unified manner. Apart from performing redundant pre-processing steps as a single instance, unified pre-processing also has one critical significance: piping of low-level inference. As shown in Figure~\ref{fig:lowinfer} certain low-level inferences such as \textit{human presence detection}, \textit{location information} etc. could serve as contextual information for a set of context-aware services~\cite{choudhury2019proactive, DASILVA20191041} and are required for their primary inference. That being said, unified pre-processing covers two key aspects: data aggregation and feature extraction.
\subsubsection{Aggregation}
To serve a service that requires a single stream from a group of sensors, individual streams from the selected sensor set are combined into a single stream. The module performs two steps as presented below.
\paragraph{Interpolation}
It must be noted that a data point might not be available in a sensor-generated stream which is needed by a service at any specific instance. This is obvious due to independent data generation period of sensors.
On a stream the interpolation can be performed as \cite{gnauck2004interpolation},
\begin{equation}
D_l = \frac{D_{t_1}-D_{t_2}}{a-b}(l-b)+D_{t_2}
\label{eq:interpolation}
\end{equation}
Here, $D_{t_1}$ and $D_{t_2}$ are two known data points in a time-series (data stream) and $a$ and $b$ are their indices, respectively. $D_l$ is an interpolated data point. The linear interpolation method stated in equation (\ref{eq:interpolation}) is effective and can be implemented easily on the stream \cite{lepot2017interpolation}. 

\paragraph{Data Fusion}
The term \textit{fusion} in this case refers to a combination of multiple data streams into a single stream. The resultant is a single continuous stream as an aggregate of the multiple input sources. 
For sensor measurements where the expected co-variance is usually low (for instance, room temperature, humidity etc.), a linear KF stated in Section \ref{kalmanfilter} has been leveraged to effectively combine the data streams into a single one. A KF is chosen over a Moving Average because, with moving average (MA), there exists challenge in the selection of the \textit{window}. A smaller window is highly affected by the noise in case of noisy measurements, whereas a larger window is much slow in reflecting the actual changes. Similarly, existing study \cite{9318810} also shows that selection of the parameter, $\alpha$, \textit{the degree of weighting decrease} ($0< \alpha\ll 1$) for Exponential moving Average (EMA), 
is a challenging task. A large value of $\alpha$ causes higher influence of noise in the output and a smaller $\alpha$ results in a slower convergence. 
\subsubsection{Feature Extraction}
The final step is to formulate feature vectors for different services which perform machine inference on sensory measurements~\cite{QIU2022241}. Recall from Eq. (\ref{eq:stream}), a data stream from a sensor is composed of data points identifiable by timestamps. 
Feature formulation is achieved through three steps.

\paragraph{Association of data and events}
The data received from heterogeneous sensory sources could be sensed independently. However, the data needs to be correlated in order to associate these measurements corresponding to a specific event and therefore to be mapped for a specific service. To achieve this, we require two information which can be exploited: \textit{time} and \textit{sensor collocation}. The timestamps assigned to the data points signify temporal correlation whereas the sensor collocation within a locality can signify spatial correlation. With that information we can define two arbitrary functions: $F_{temp}()$ and $F_{spat}()$. The outcome of these two functions are binary and signify if there exists a temporal and spatial correlation within the set of data points and need not be individually defined for each stakeholder services. Therefore, the necessary set of data points for a specific event and a specific service can be isolated with a simple logic as,

\begin{equation}
\text{Association exist }=
    \begin{cases}
      1 & \text{if} (F_{temp}() \:AND \:F_{spat}())\\
      0 & \text{otherwise}\\
    \end{cases}       
\end{equation}
The associated set of data points signify multimodal assessment corresponding to a specific event.

\paragraph{Dynamic window formulation}
The next task is to formulate window on the data streams for feature extraction. Let $t_i$ be the timestamp at which a window is to be formed on correlated data streams for the application $i$. Let, $W_i$ be the size of the window required. In that case the window can be represented as, $W_i=t_B-t_A$ where, $t_B$ and $t_A$ are the timestamps at the two end points of the window
\begin{equation}
\begin{array}{l}
    t_A=t_i-k_l\times\delta t\\
    t_B=t_i+k_r\times\delta t
\end{array}
\end{equation}
where, $k_l$ and $k_r$ are the no. of intervals of size $\delta_t$ on either side of $t_i$ along the time axis.
\paragraph{Feature vector formulation}
A multi-dimensional feature vector is composed of information from heterogeneous sensors. Each stake-holder service requires specific constituents for its feature vector. For instance, an HVAC and an ambience lighting system may both require occupancy estimate for feature formulation. In such a case, the occupancy estimate, which is derived from a set of sensory measurements is computed once and can be utilized by all the available applications towards an optimal resource utilization. Let $F$ be a set of functions which generate individual features from the sensory data. Therefore, $F=\{F_1(), F_2(),...F_n()\}$ is a finite set of arbitrary functions for feature extraction.  
\section{Experimental setup}
\begin{figure}[tb]
\centering
\includegraphics[width=1\columnwidth]{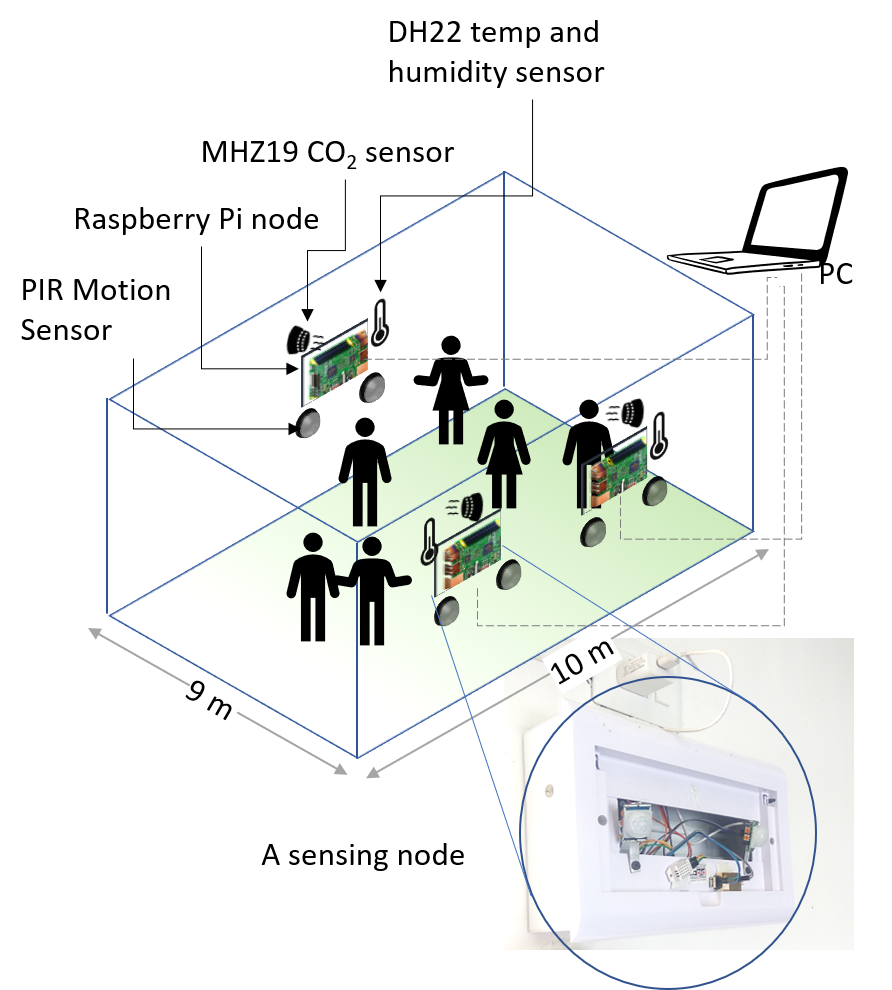}
\caption{Experimental setup}
\label{fig:room2}
\end{figure}

To analyze the performance of \ourmethod we deploy a room-scale testbed, in our institute campus as shown in Figure \ref{fig:room2}. 
The testbed employs 3 nodes, each having a temperature sensor, a humidity sensor, a CO\textsubscript{2} sensor and 2 instances of motion sensors. Due to the presence of multiple instances, redundant data is generated. But, the service requires one value each from each types of sensors. Adhering to the acquisition limitations, temperature and humidity sensors were set to generate data at an interval of 3 seconds, the CO\textsubscript{2} sensors, an interval of 30 seconds and the motion sensors have 1 second each. The sensors generate data in the form of streams. Each of these streams is fed to a PC (8-core i7 3.6GHz, RAM: 8GB) where \ourmethod performs the necessary computation. We consider a service for detecting the room occupancy~\cite{candanedo2016accurate} based on ambient sensing, where multiple sensors are fused to estimate the occupancy level. Each feature representation contains one instance of temperature and humidity measurements and the set of motion sensor measurements. We consider another service for real-time indoor air-quality monitoring~\cite{benammar2018modular} based on temperature, humidity and CO\textsubscript{2} in order to study a multi-service scenario. 

We evaluate the performance of \ourmethod primarily from two aspects:

(I) the performance evaluation of the sensor selection approach and 

(II) the resource utilization of the shared infrastructure.

For evaluating the sensor selection approach, we first study, what is its impact on the performance of inference on a service. For this, we compare \ourmethod with TOPSIS approach for sensor selection~\cite{nunes2018elimination} which unlike \ourmethod is not adaptive to the changing dynamics of the sensory system .
\begin{figure}[tb]
            \centering
            \includegraphics[width=0.8\columnwidth]{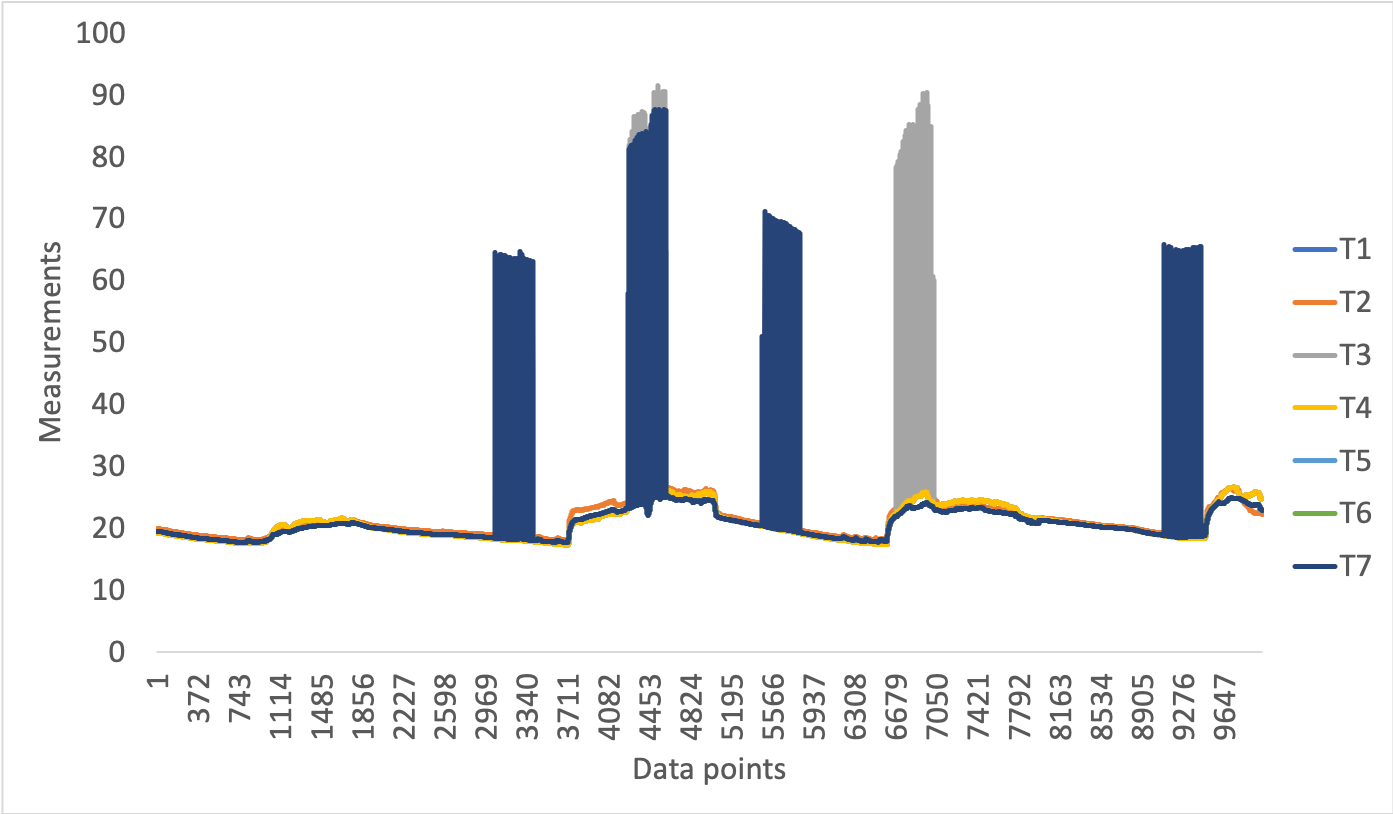}
            \caption{\small 7 Data sources}
            \label{fig:originaldata}
\end{figure}
Since, we do not have ground truth information about the optimal sensor given a selection criteria, we rather evaluate the impact of the proposed adaptive sensor selection approach by studying its impact on a stakeholder service. For that a knowledge file trained with 2000 instances of features is employed in a service for \textit{occupancy detection} say $Service_A$. We use different model to develop $Service_A$ viz. a K-Nearest Neighbor (KNN), a Logical Regression (LR), a Decision Tree (DT) and a Support Vector Machine (SVM)  to establish the correctness of \ourmethod.  As for continuous monitoring, $Service_A$ is invoked periodically in three scenarios: i) when the sensor selected by \ourmethod is used to select a sensor from a given set and ii) the sensory measurements from the available instances were aggregated during the feature formulation without sensor selection and iii) when a sensor is selected by TOPSIS. Since, the primary objective of incorporation of the proposed sensor selection approach is to reduce the resource consumption footprint, therefore we show how \ourmethod can effect the computation time of $Service_A$. We also compare the amount of data being processed with optimal sensor is selection for $Service_A$. The availability of sensor data which have measurement incorrectness is pretty scarce. Therefore, to study the QoS awareness, where we take random error into account, we make use of the benchmark dataset prepared in~\cite{de2016benchmark} which contains random measurement incorrectness exhibited by \textit{temperature sensors}. From this dataset, we address a scenario as shown in Figure ~\ref{fig:originaldata} where, 7 instances of temperature sensors ($Tn$) are in play. As per the documentation of the dataset, here, $T3$ and $T7$ generated incorrect measurements at various instances. 
For the evaluation, first we study, how the sensor selection was achieved when sensor accuracy was preferred and then we prioritize the sensor reliability to select the most optimal sensor.

\begin{figure*}[t]
\vspace{-9pt}
        \centering
        \begin{subfigure}[b]{0.405\textwidth}
            \centering
            \includegraphics[width=\textwidth]{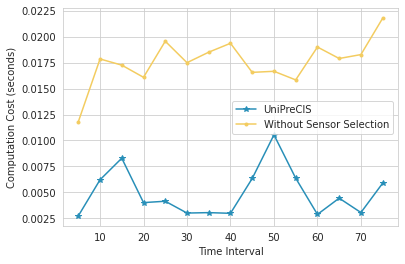}
            \caption[Network2]%
            {{\small Using SVM}}
            \label{fig:phasewithoutpeople}
        \end{subfigure}
        \hfill
        \begin{subfigure}[b]{0.405\textwidth}   \centering 
            \includegraphics[width=\textwidth]{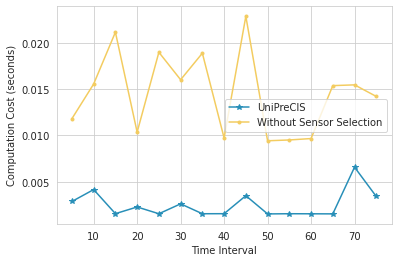}
            \caption[]%
            {{\small Using LR}}   
            \label{fig:phasewithpeople}
        \end{subfigure}
        \begin{subfigure}[b]{0.405\textwidth}   \centering 
            \includegraphics[width=\textwidth]{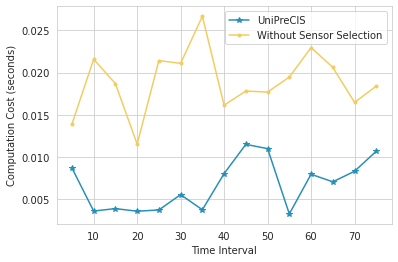}
            \caption[]%
            {{\small Using KNN}}  
            \label{fig:magwithpeople}
        \end{subfigure}
        \hfill
        \hfill
        \begin{subfigure}[b]{0.405\textwidth} \centering 
            \includegraphics[width=\textwidth]{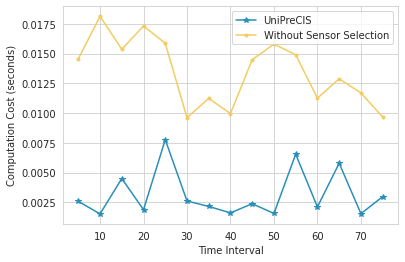}
            \caption[]%
            {{\small Using DT}}    
            \label{fig:magwithoutpeople}
        \end{subfigure}
        \caption[ The response time of the   ]
        {\small The response time of the service while predicting the occupancy in R1 every 5 minutes} 
        \label{fig:r1response}
   \end{figure*}
To evaluate the proposed pre-processing component, we study the computation duration and memory utilization footprint of the running services with and without the unified pre-processing. Here, the occupancy detection service is invoked at an interval of 5 seconds and the air quality measurement service is invoked every second. Therefore we have a scenario of parallel execution of services that use features with different data granularity and also depend on a common set of sensors in a collocated fashion. When the framework is not used, each service perform their own set of pre-processing and respective inferences independently. Whereas, when \ourmethod is in use, it performs the pre-processing steps in an unified manner and provides the depending services with their individual feature vectors. 
\begin{figure}[tb]
\centering
\includegraphics[width=0.9\columnwidth]{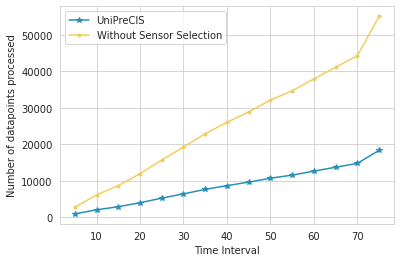}
\caption{Comparison of amount of data points processed}
\label{fig:amount}
\end{figure}
\begin{figure}[tb]
\vspace{-9pt}
        \centering
        \begin{subfigure}[b]{0.44\textwidth}
            \centering
            \includegraphics[width=\textwidth]{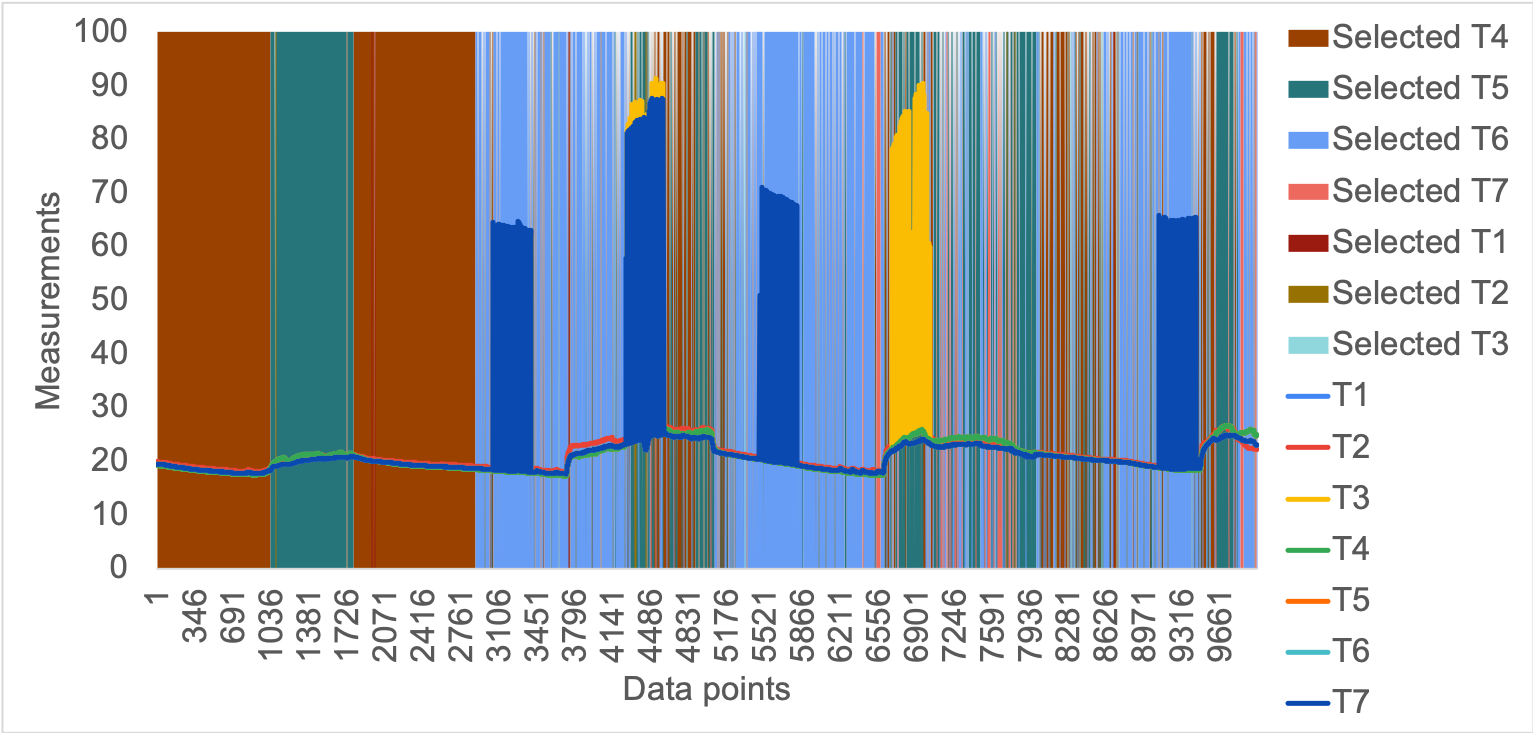}
            \caption[Network2]%
            {{\small Selected sensors}}
            \label{fig:noaccuracy}
        \end{subfigure}
        \hfill
        \hfill
        \begin{subfigure}[b]{0.44\textwidth}   
            \centering 
            \includegraphics[width=\textwidth]{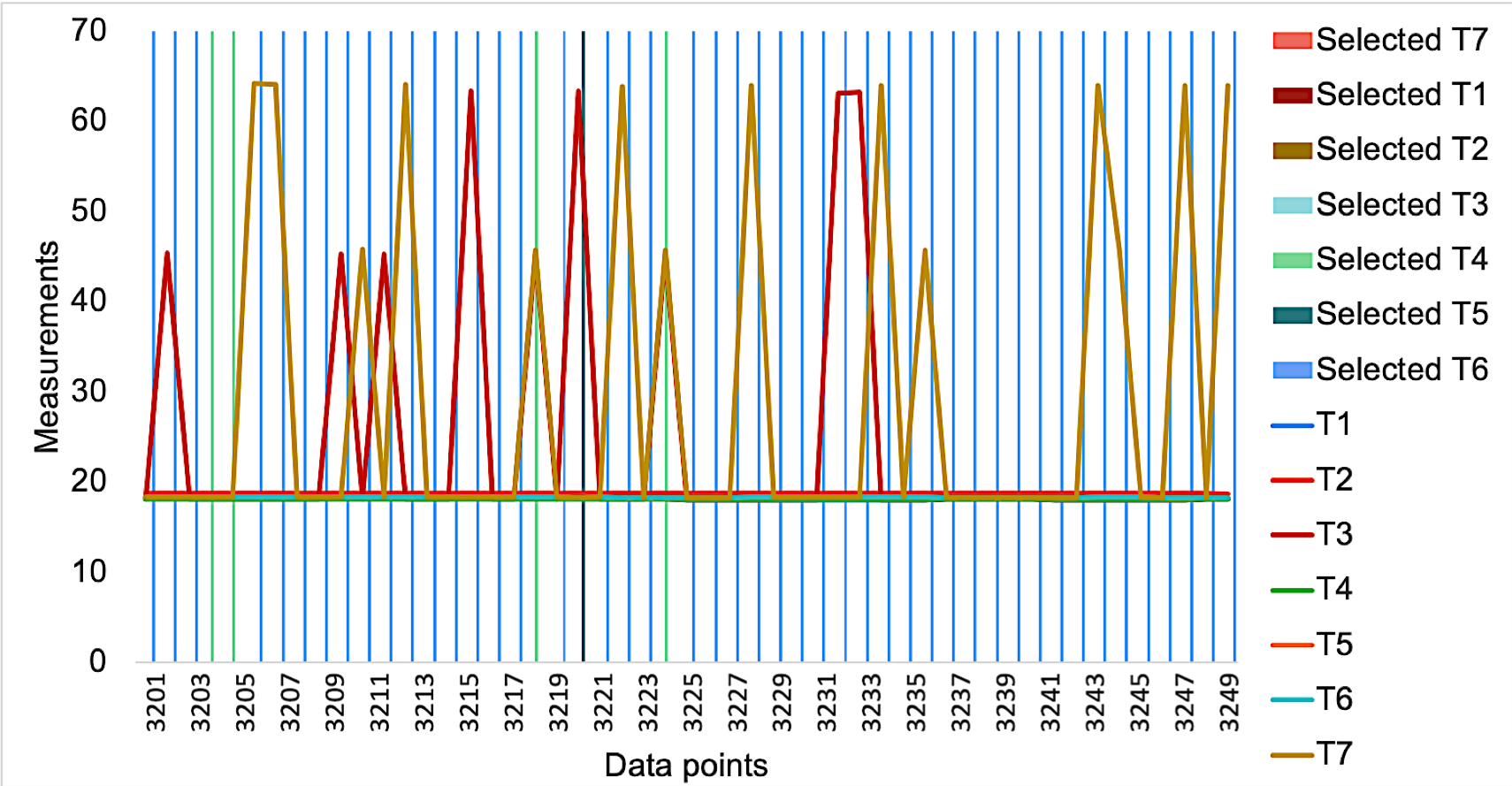}
            \caption[]%
            {{\small Close-up view for a better visibility}}    
            \label{fig:withaccuracy1}
        \end{subfigure}
        \caption[ The response time of the   ]
        {\small Sensor selection applied prioritizing accuracy} 
        \label{fig:selectionaccuracy}
   \end{figure}
\begin{figure}[!ht]
\vspace{-9pt}
        \centering
        \begin{subfigure}[b]{0.44\textwidth}
            \centering
            \includegraphics[width=\textwidth]{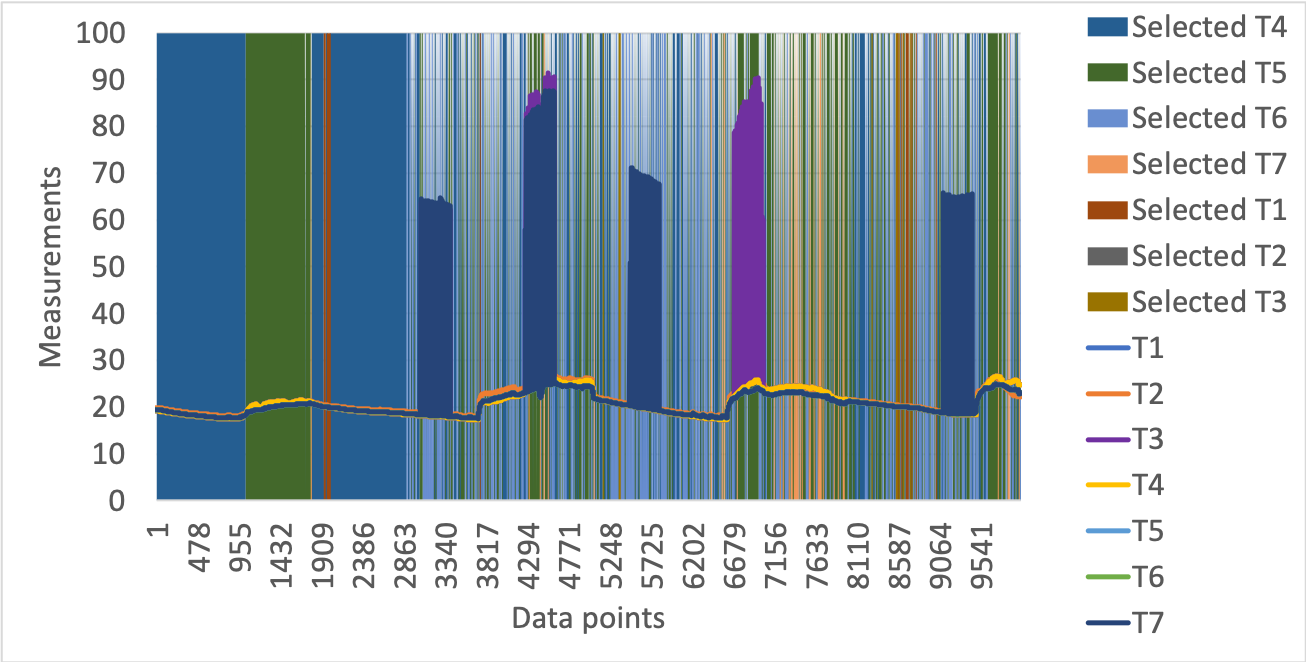}
            \caption[Network2]%
            {{\small Selected sensors}}
            \label{fig:noaccuracy12}
        \end{subfigure}
        \hfill
        \hfill
        \begin{subfigure}[b]{0.44\textwidth}   
            \centering 
            \includegraphics[width=\textwidth]{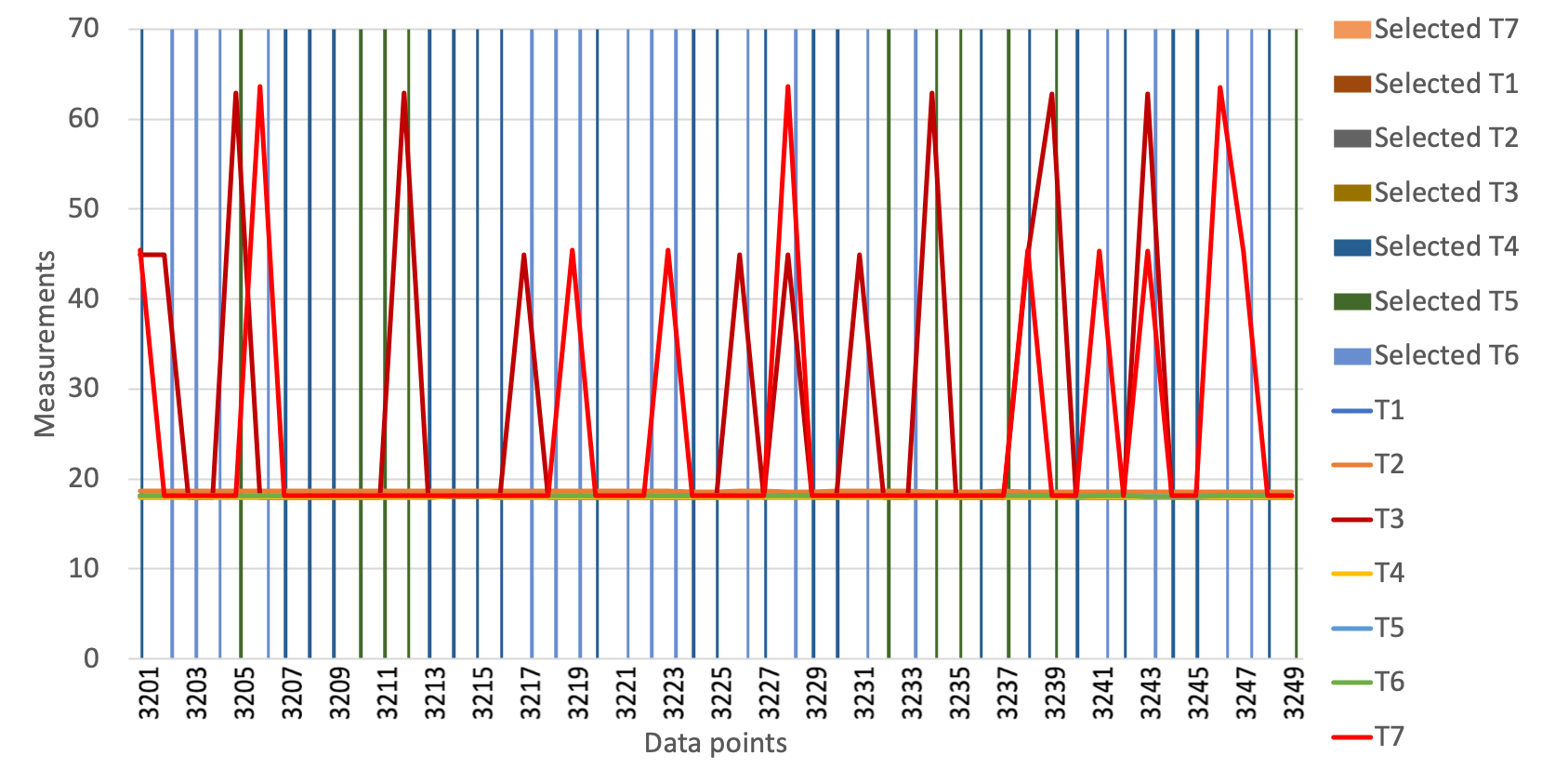}
            \caption[]%
            {{\small Close-up view for a better visibility}}    
            \label{fig:withaccuracy12}
        \end{subfigure}
        \caption[ The response time of the   ]
        {\small Sensor selection applied prioritizing reliability} 
        \label{fig:selectionaccuracy2}
   \end{figure}
 \begin{figure}[htb]
\centering
\includegraphics[scale=0.33, keepaspectratio]{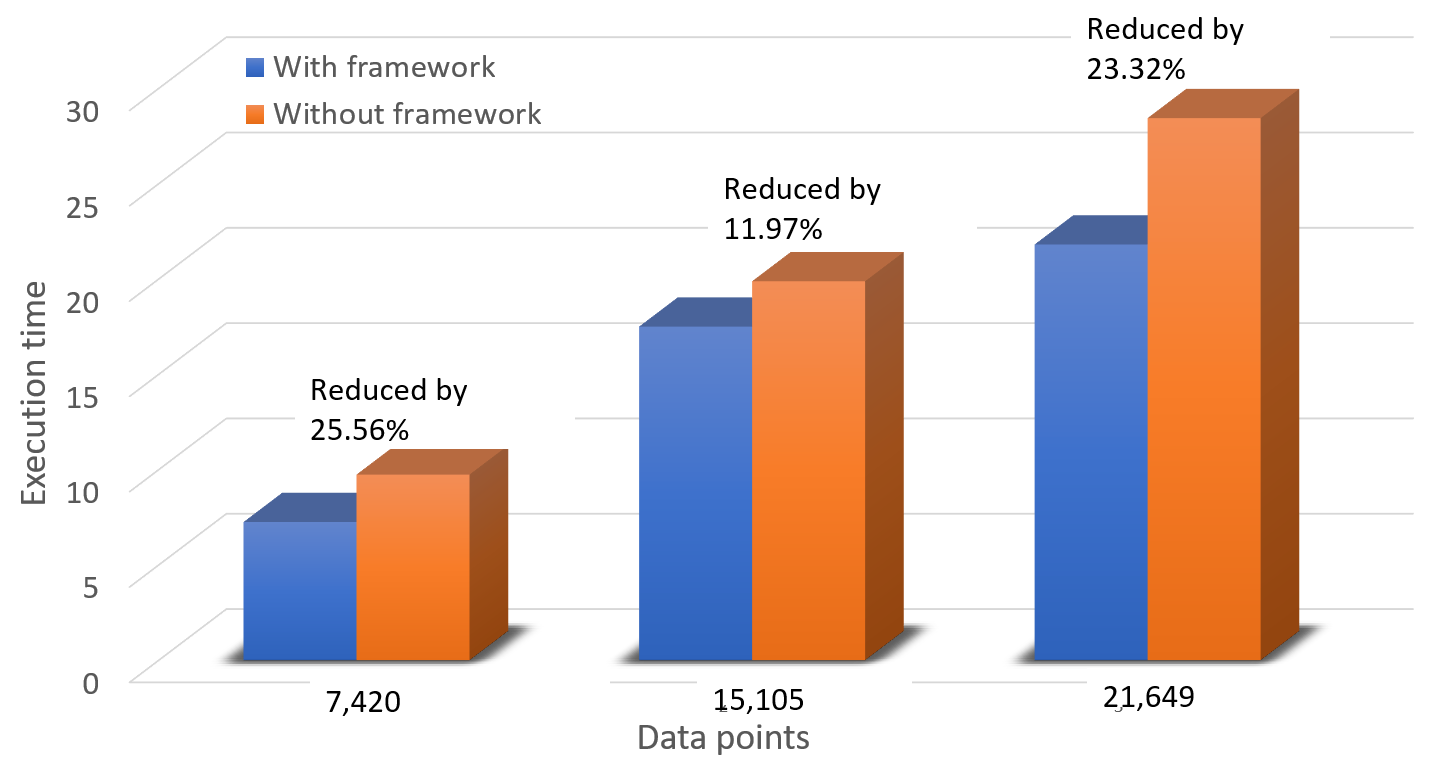}
\caption{Impact on overall execution time}
\label{fig:execution-total}
\end{figure}
\begin{figure*}[htb]
\vspace{-9pt}
        \centering
        \begin{subfigure}[b]{0.285\textwidth}
            \centering
            \includegraphics[width=\columnwidth]{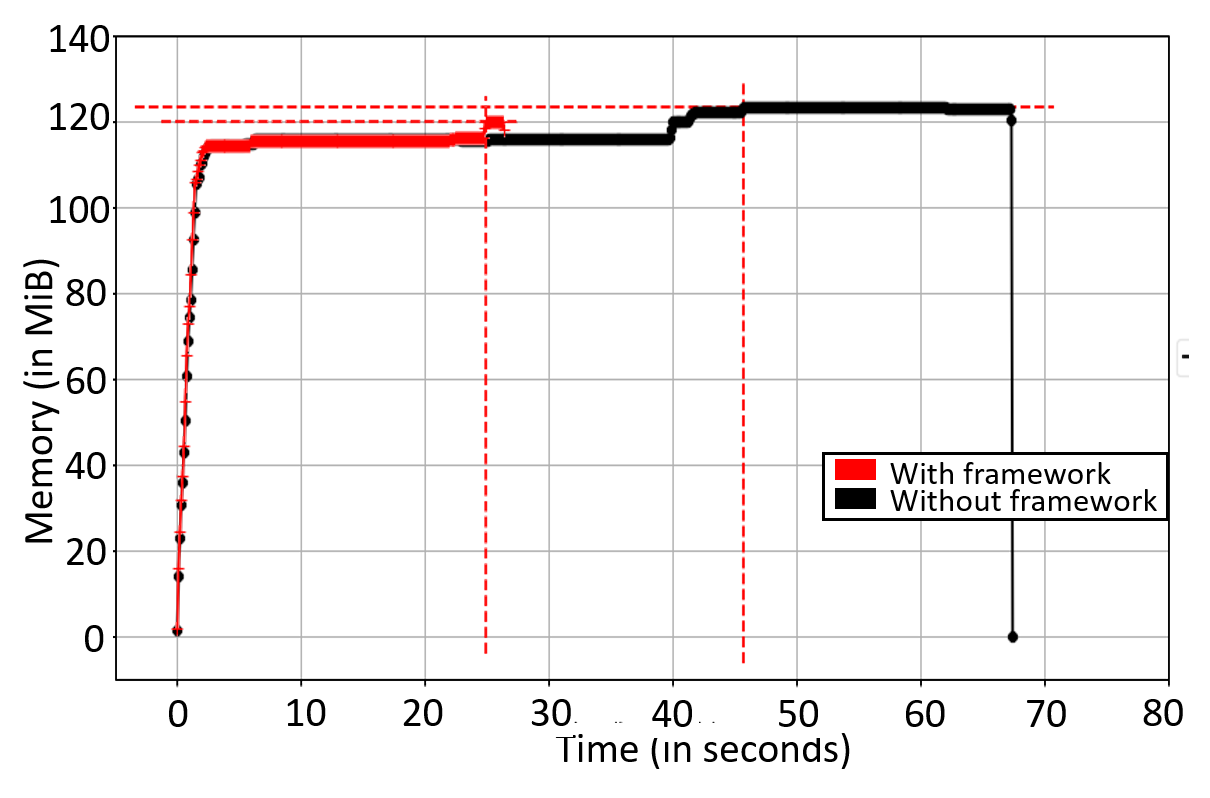}
            \caption[Network2]%
            {{\small Data of 7,420 seconds}}
            \label{fig:point1}
        \end{subfigure}
        \begin{subfigure}[b]{0.335\textwidth}   
            \centering 
            \includegraphics[width=\columnwidth]{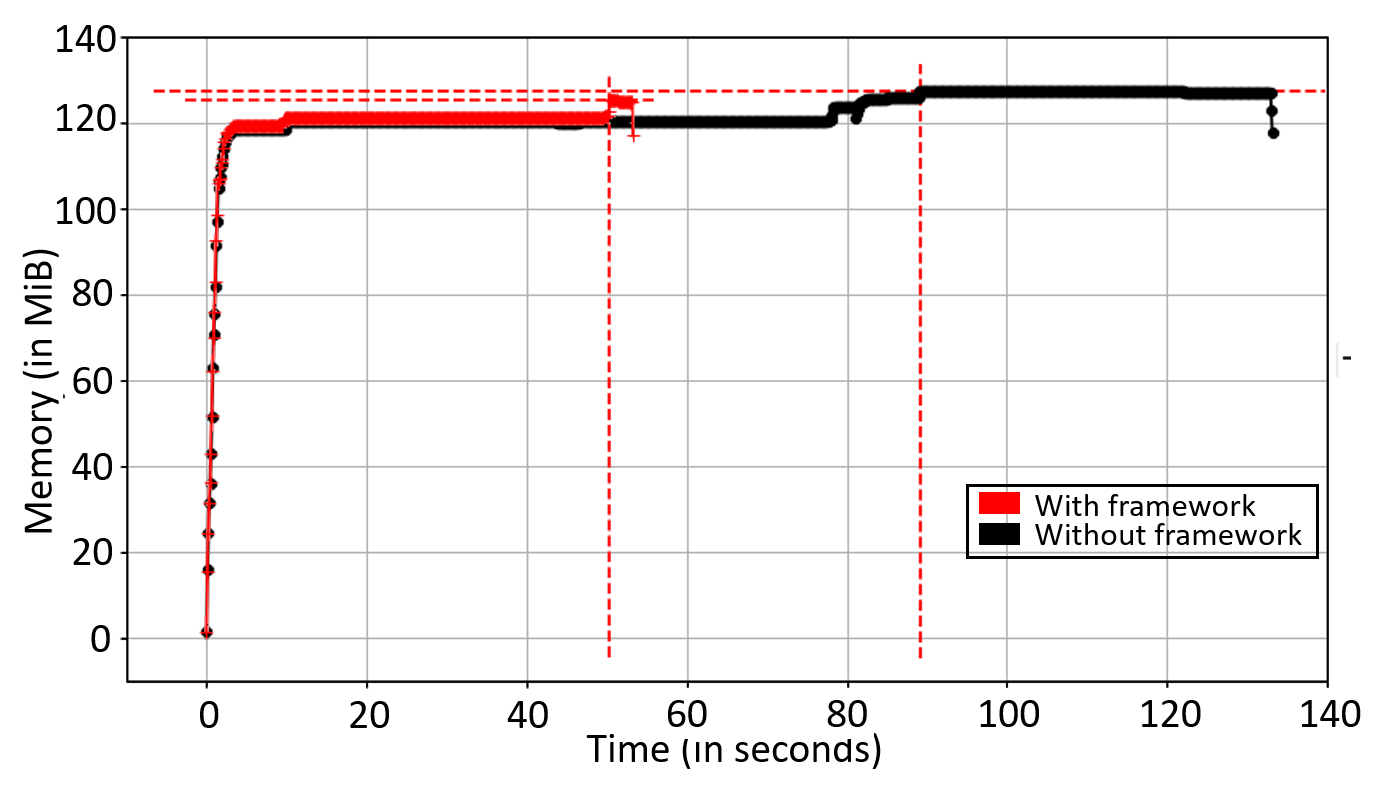}
            \caption[]%
            {{\small Data of 15,105 seconds}}   
            \label{fig:point2}
        \end{subfigure}
        \begin{subfigure}[b]{0.365\textwidth}   
            \centering 
            \includegraphics[width=\columnwidth]{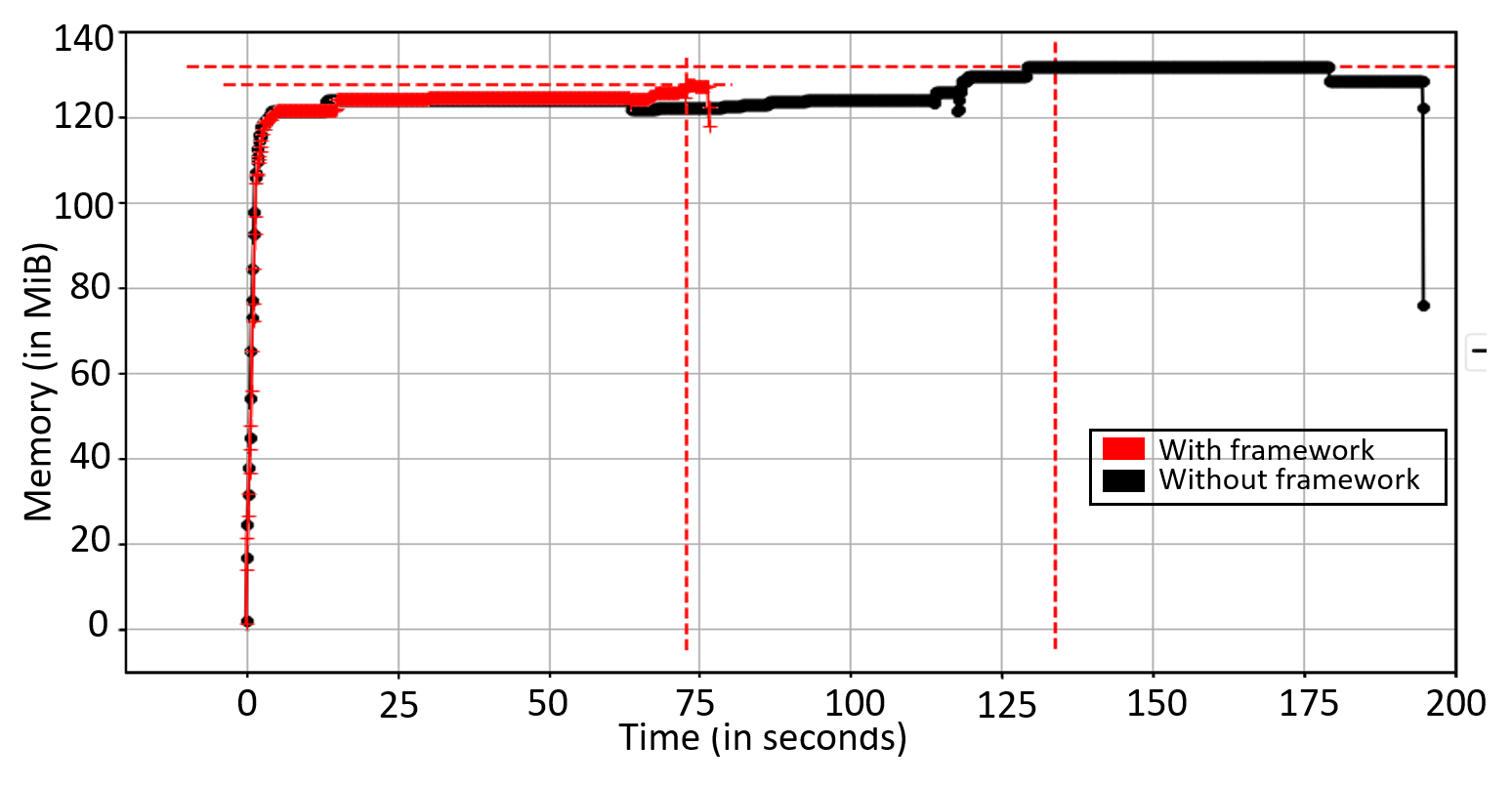}
            \caption[]%
            {{\small Data of 21,649 seconds}}   
            \label{fig:point2}
        \end{subfigure}
        \caption[ The response time of the   ]
        {\small Comparison of Memory consumption profile by the services when the minimum granularity of data is kept as 1 sec interval} 
        \label{fig:memprofile}
   \end{figure*}

\section{Results and analysis}
Figure \ref{fig:r1response} shows the processing time of the service $Service_A$, while being invoked at a regular interval (5 minutes in this case). It must be noted that the aim of the sensor selection mechanism is to reduce the data to be processed while maintaining the service performance optimal.
It is observable from Table \ref{tab:accuracy2} the prediction accuracy with selection could reach upto 90\% but never less than the counterpart. Figure \ref{fig:amount} presents the amount of data points processed over time. It clearly shows that sensor selection induced a reduced amount of data processing, thereby preserving valuable computation resources. Next, we invoke sensor selection on the data from~\cite{de2016benchmark}
as shown in Figure \ref{fig:originaldata}). With accuracy as priority, we observe that the proposed approach does not select the sensors $T3$ or $T7$ whenever they were generating spurious values as shown in Figure \ref{fig:selectionaccuracy}. Whereas, as we prioritize reliability with $t=100$ (similar results were obtained at both $t=50$ and $t=30$) the proposed approach avoids $T3$ and $T7$ due to least reliability factors and is shown in Figure \ref{fig:selectionaccuracy2}. Since, TOPSIS~\cite{nunes2018elimination} does not consider the dynamic change in sensor characteristics, therefore, with the requirement specification being constant TOPSIS will not change the sensor selected over the course of time. Hence, we avoid showing any outcome from TOPSIS. 
Moving to the performance of the proposed pre-processing component, for the two available services, we observe the overhead reduction in terms of execution time to upto 25\% for 3 different duration of generated data in Figure \ref{fig:execution-total}. The corresponding 3 cases of memory consumption profile is shown in Figure \ref{fig:memprofile}. The memory profile clearly shows upto 10\% reduction in maximum memory footprint by \ourmethod.

\begin{table}[h]
\caption{Accuracy demonstrated on an average by the classifiers with and without sensor selection}
\label{tab:accuracy2}
\resizebox{\columnwidth}{!}{%
\begin{tabular}{
>{\columncolor[HTML]{EFEFEF}}l lll}
Classifier                                        & \cellcolor[HTML]{EFEFEF}\begin{tabular}[c]{@{}l@{}}Without sensor\\ selection (fused)\end{tabular} & \cellcolor[HTML]{EFEFEF}TOPSIS & \cellcolor[HTML]{EFEFEF}UniPreCIS \\ \hline
\multicolumn{1}{|l|}{\cellcolor[HTML]{EFEFEF}KNN} & \multicolumn{1}{l|}{53.6}                                                                          & \multicolumn{1}{l|}{39.96}     & \multicolumn{1}{l|}{86.44}        \\ \hline
\multicolumn{1}{|l|}{\cellcolor[HTML]{EFEFEF}DT}  & \multicolumn{1}{l|}{36.59}                                                                         & \multicolumn{1}{l|}{39.73}     & \multicolumn{1}{l|}{84.21}        \\ \hline
\multicolumn{1}{|l|}{\cellcolor[HTML]{EFEFEF}LR}  & \multicolumn{1}{l|}{73.7}                                                                          & \multicolumn{1}{l|}{38.83}     & \multicolumn{1}{l|}{84.2}         \\ \hline
\multicolumn{1}{|l|}{\cellcolor[HTML]{EFEFEF}SVM} & \multicolumn{1}{l|}{56.08}                                                                         & \multicolumn{1}{l|}{42.99}     & \multicolumn{1}{l|}{94.31}        \\ \hline
\end{tabular}}
\end{table}

\section{Literature Survey} 
We aim to perform a unified data pre-processing and sensor ranking \& selection in order to shape the data in terms of quality and quantity. 
We present the relevant state-of-the-art works in the following subsections, focusing mainly on data pre-processing and sensor selection approaches.
\subsection{Data Pre-processing for Multi-sensor Data Fusion}
Multi-sensor data fusion \cite{XIAO201923} is a quite prominent research topic in the domains of sensory bigdata analytics and has observed a plethora of usecases over the last decades. The idea is to combine data from multiple sensors for a more specific inference \cite{liggins2017handbook}. The pre-processing on the data involves various activities including the reduction of instances~\cite{ramirez2017survey}.
Towards this pre-processing, very few works \cite{kenda2019streaming, yin2020novel} deal with the streaming data specific to IoT environment. And it is difficult to find a generalized approach. In a multi-service scenario there exist a need to bridge the gap between data accumulation and feature formulation for the IoT applications~\cite{peros2021ermis}. A set of pre-processing steps were studied by the earlier works for big-data~\cite{garcia2016big} which highlights the significant steps involved. The authors in \cite{kenda2019streaming}, propose a generalized architecture for pre-processing involving stream data fusion. This is a significant motivation towards our proposed pre-processing mechanism that is directed towards a shared IoT infrastructure and introduces mapping of resources to multiple stake-holder services. However, their work does not exploit a collocated multi-service scenario and mainly focuses on applications which use incremental learning. Moreover, their framework does not incorporate measures for QoS aware sensory sources. 
\subsection{Sensor Ranking and Selection for Optimization}
To select a set of most optimal sensors among an available set has been of interest for quite some time. In the literature, many studies~\cite{costa2021goat, saito2021determinant} can be seen for sensor ranking and selection. The primary goal is to assign some rank to the available sensors based on certain factors. The ranking can be done based on the remaining energy, relevance towards the task in hand, location, etc. It can be observed that most of the works of ranking are domain specific~\cite{bharti2019value, bharti2020optimal}. Additionally, the majority~\cite{babu2019context, perera2013context} consider the sensor characteristics in a static fashion. It's a major setback towards making a system adaptive. For instance, in \cite{perera2013context}, the authors consider sensor characteristics such as accuracy, range of measurement, precision, latency etc. for selecting an optimal sensor. The modelling is performed by considering the measurements as discrete events, which fails to encompass the longitudinal behaviour of the sensors. In this context to make the selection adaptive, we model two sensor characteristics: \textit{Accuracy} and \textit{Reliability}. The relevant literature studies are stated below:
\subsubsection{Modelling of Accuracy}
The sensor systems are susceptible to inaccurate measurements due to various underlying factors. Even the smartphone sensors, one of the most pervasive kind of IoT devices, vary in measurement accuracy \cite{kuhlmann2021smartphone}. Existing approaches for accuracy estimation can be categorized into three main classes: \textit{voting based, inference based and learning based}.
The voting based approaches primarily consider measurements from multiple homogeneous sensors and take a collective decision. Thus, it is a perfect fit when there is collocation. It has been used across various domains \cite{tsarev2019fuzzy} addressing variety of problems \cite{tsarev2018classification}. The voting based approaches enjoy a significant advantage over all the others as the requirement of any \textit{priors} is minimal or in most of the cases, nil. The inference based approaches estimate the true state of a sensor measurement based on a set of reported measurements and/or a set of priors. The static inference is when the estimation is performed based on current measurements and the priors, whereas, the dynamic inference considers a set of previous states of measurements. The learning based approaches \cite{yan2020collaborative} need a model to be trained with data. A static variant requires the training to be performed once and a dynamic variant keeps updating the knowledge with new observations. The most recent studies prefer machine learning for estimating the accuracy of sensors. But it has its limitations due to the requirement of domain knowledge. Also, the problem is usually formulated to address the accuracy of the model \cite{ozcan2020accurate} rather than accuracy of the sensor device itself. 
An earlier study \cite{wen2014accuracy} on performance evaluation of these approaches clearly shows that the learning based approaches are better among all but at a significantly high computation cost. The findings argue that the dynamic inference based approaches (such as KF) are comparable with learning based approaches with a significantly lower computation cost. 
That said, we incorporate the high performance of dynamic inference based approach at low cost with the advantage of voting based approach and derive a novel estimation strategy. 
\subsubsection{Modelling of Reliability}
\label{sec:reliability}
\begin{figure}[ht]
\centering
\includegraphics[scale=.24, keepaspectratio]{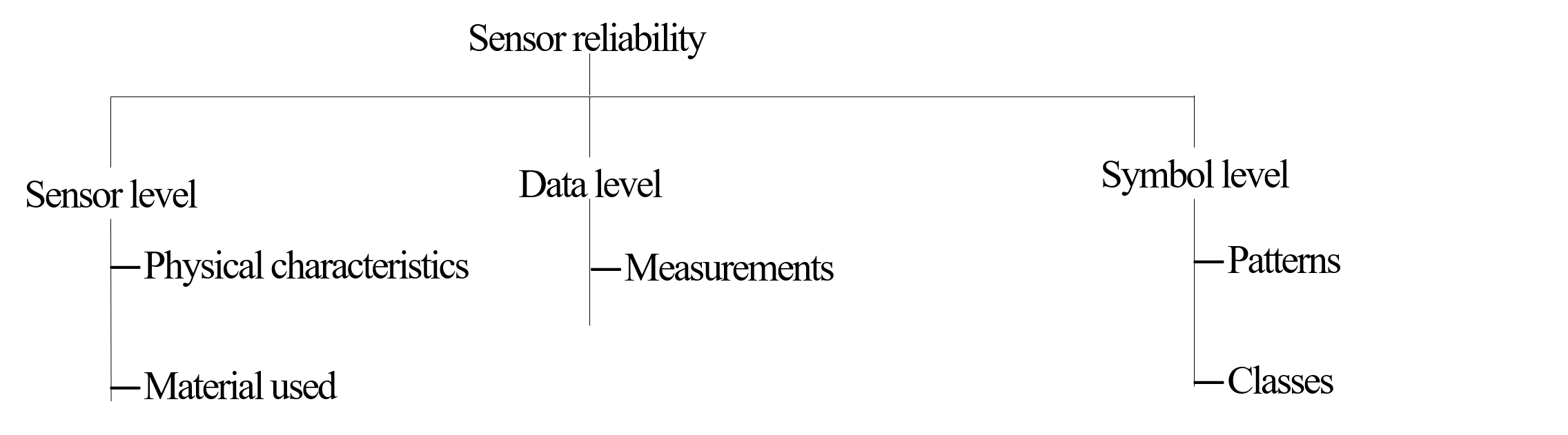}
\caption{Problem of reliability addressed at different levels}
\label{fig:reliabilitytree}
\end{figure}

Sensors can report incorrect readings because of the dynamically changing operating conditions such as temperature, humidity, etc. or due to bias and calibration drifts caused by age related degradation. Even after performing sensing and \textit{quantization}, the data can still be affected by various hardware errors such as \textit{crosstalk} and radiation effects \cite{mukhopadhyay2008model}. These errors can differ widely in terms of severity, frequency of occurrence, and other statistical properties. In literature, the reliability of a sensor has been addressed at various levels as shown in Figure\ref{fig:reliabilitytree} 

We consider the problem of reliability in a data oriented manner. 
This can be addressed in terms of \textit{the extent of inconsistency in reported data by a sensor with respect to the phenomenon of interest’s true behavior} \cite{ni2009sensor}. The state of the art approaches for modelling this see mostly machine learning based approaches \cite{noshad2019fault, zidi2017fault}. Although reported performances are notable, they suffer from a common disadvantage, \textit{application dependency}. Therefore, for a heterogeneous environment, different models need to be trained in order to circumvent available objectives. The machine based training and inference are usually computationally expensive and require suitable datasets for effective performance. In addition to that, the available datasets, in almost all cases were prepared based on simulated parameters \cite{de2016benchmark}. The other significant approaches are based on evidence theory \cite{ghosh2020fault, jiwei2018reliability}. But, addressing reliability following this approach is still at theoretical level. The analysis of the study is usually performed based on several assumptions for parameters such as \textit{masses}, whose practical relevance is questionable. 
With the availability of multi-instances of sensors along with existing collocations, low computation mathematical models can be a good fit for the problem of reliability.

\section{Conclusion}
As IoT infrastructure is becoming part of our daily life, more and more smart applications are emerging on regular basis. A single infrastructure can have numerous such applications running as services on a shared setup. In such a scenario, the services can be decoupled from the hardware infrastructure to enhance cost effective deployment. But, leveraging such existing infrastructure introduce new challenges while supporting the QoS factors of these services. That's where we introduce \ourmethod which facilitates optimal sensor selection addressing the QoS requirements of the stakeholder services. Apart from isolating optimal sensors, \ourmethod also performs unified pre-processing of the sensory data for collocated services thereby minimizing redundant computation by individual parts. A testbed study along with dataset based validation shows that \ourmethod can adapt to the changing dynamics of the setup while selecting the most optimal set of sensors. At the same time it preserves the performance of the depending applications by maintaining the inference accuracy. The unified pre-processing also shows a reduction in the computation time by upto 25\%. In this work, the dynamicity of two sensor characteristics viz. \textit{accuracy and reliability} has been modeled. However, modelling more such sensor properties remains to be explored. Additionally, \ourmethod has been evaluated only on two possible smart services however, the performance of this while considering more such services remains to be explored in our future work.

\section{Appendix I}
\label{appendix}
Proof that $C_{x_i}$ is optimal:
Let, X be the set of all optimal values given a set of values, $V$, $X \subseteq V$

Case 1) When all the values of $x_i$ are equal.
($\therefore x_1 = x_2 = x_3= … =x_n$)
Then,  
\begin{equation}
\begin{aligned}
    & \;\;\;\;\;\; x_i-x_j=0,\ \forall\ i \ and \ j \\
    & \Rightarrow (x_i-x_j)^p=0 \\
    & \Rightarrow e^{-\frac{(x_i-x_j)^p}{c}}=e^0=1=U(x_i, x_j), \ \forall\ i \ and \ j \\
    & \Rightarrow \sum_{i=1}^{n} U(x_i, x_j)=n\\
    & \Rightarrow C_{x_i}=n\\
\end{aligned}
\end{equation}
Thus, in this case, all $C_{x_i}$  are equal and all values are optimal.

Case 2) When all the values of $x_i$ can be arranged in ascending order and are of equal difference.

$\therefore$ the values are in AP

$\therefore$ For a given value of $x_i$ , $(x_i-x_j)^p$ will give two APs  for different values of $j$ (one for $i<j$ and another for $i>j$)

$\therefore$ For a given value of $x_i$, $e^{-\frac{{(x_i-x_j)}^p}{c}}$ will be in two GPs and the sum of these two GPs is a constant.

$\therefore$ The values of $U(x_i, x_j)$ is the same for any specific value of $i$.

Case 3) All other cases,

Let $k$ be no. of elements in the set of optimal values, $X$.
$\therefore$ Given, $x_i, x_j \in X$, $(x_i-x_j)^p$ will be less than or equal to $(x_i-x_m)^p$, $\forall$ $x_i, x_j \in X$ and $x_m \notin X$ considering the linear distance between each pair of measurements.
\begin{equation}
\begin{aligned}
    & \therefore (x_i-x_j)^p\leq(x_i-x_m)^p,\ \forall\ x_i, x_j \in X \ and \ x_m \notin X \\
    & \Rightarrow -(x_i-x_j)^p\geq-(x_i-x_m)^p \\
    & \Rightarrow e^{-\frac{(x_i-x_j)^p}{c}}\geq   e^{-\frac{(x_i-x_m)^p}{c}}\\
    & \Rightarrow \sum_{j=1, x_i\in X}^{n} e^{-\frac{(x_i-x_j)^p}{c}}\geq\sum_{m=1, x_m\notin X}^{n} e^{-\frac{(x_i-x_m)^p}{c}}\\
    & \Rightarrow \underbrace{C_{x_i}}_{x_i\in X} \geq \underbrace{C_{x_i}}_{x_i\notin X}\\
\end{aligned}
\end{equation}
Example: Let the values measured be, $x_1=0.6$, $x_2=12.0$, $x_3=11.8$, $x_4=11.9$, $x_5=10.0$

On cognitive analysis, we can say that the near optimal measurements lie among, $x_2, x_3, x_4$ and $x_5$ as $x_1$ is too far a value from the rest. 

\[ \text{After normalization, using,\:\:\:}x_i = \frac{x_i - \overline{x}}{\delta}\]
we get,
${x_1=-2.16, x_2=0.68, x_3=0.63, x_4=0.66, x_5=0.18}$
Now we formulate $M$ using membership function, $f(.)$,
\[f(x_i, x_j)=e^{-\frac{{(x_i-x_j)}^2}{2}}\]
\begin{center}
\begin{table}[htb]
\label{tab:my-table}
\begin{tabular}{llllll}
                        & $X_1$                     & $X_2$                     & $X_3$                     & $X_4$                     & $X_5$                     \\ \cline{2-6} 
\multicolumn{1}{l|}{$X_1$} & \multicolumn{1}{l|}{1} & \multicolumn{1}{l|}{-} & \multicolumn{1}{l|}{-} & \multicolumn{1}{l|}{-} & \multicolumn{1}{l|}{-} \\
\multicolumn{1}{l|}{$X_2$} & \multicolumn{1}{l|}{0.018} & \multicolumn{1}{l|}{1} & \multicolumn{1}{l|}{-} & \multicolumn{1}{l|}{-} & \multicolumn{1}{l|}{-} \\
\multicolumn{1}{l|}{$X_3$} & \multicolumn{1}{l|}{0.02}  & \multicolumn{1}{l|}{0.999}  & \multicolumn{1}{l|}{1}  & \multicolumn{1}{l|}{-}  & \multicolumn{1}{l|}{-}  \\
\multicolumn{1}{l|}{$X_4$} & \multicolumn{1}{l|}{0.019}  & \multicolumn{1}{l|}{0.999}  & \multicolumn{1}{l|}{0.999}  & \multicolumn{1}{l|}{1}  & \multicolumn{1}{l|}{-}  \\
\multicolumn{1}{l|}{$X_5$} & \multicolumn{1}{l|}{0.064}  & \multicolumn{1}{l|}{0.883}  & \multicolumn{1}{l|}{0.90}  & \multicolumn{1}{l|}{0.894}  & \multicolumn{1}{l|}{1}  \\ \cline{2-6} 
\end{tabular}%
\end{table}
\end{center}
Therefore, we have the values of $C_{x_i}$'s
\begin{center}
\begin{tabular}{|c|c|c|c|c|c|}
\centering
$C_{x_1 }$&	$C_{x_2 }$	&$C_{x_3 }$&	$C_{x_4 }$&	$C_{x_5}$\\
\hline
1.121&	3.899&	3.918&	3.911&	2.741
\end{tabular}
\end{center}

In sorted order,\:\:
\begin{tabular}{|c|c|c|c|c|c|}
\centering
$C_{x_3 }$&	$C_{x_4 }$	&$C_{x_2 }$&	$C_{x_5 }$&	$C_{x_1}$\\
\hline
3.918&	3.911&	3.899&	2.741&	1.121
\end{tabular}
Here, if $k=3, (k>n/2)$ then the selection becomes $X=\{x_4, x_3, x_2\}$ matching our cognitive assumption.

\medskip

\bibliography{mybibfile}

\begin{thebibliography}{10}
\expandafter\ifx\csname url\endcsname\relax
  \def\url#1{\texttt{#1}}\fi
\expandafter\ifx\csname urlprefix\endcsname\relax\def\urlprefix{URL }\fi
\expandafter\ifx\csname href\endcsname\relax
  \def\href#1#2{#2} \def\path#1{#1}\fi

\bibitem{ranieri2021activity}
C.~M. Ranieri, S.~MacLeod, M.~Dragone, P.~A. Vargas, R.~A.~F. Romero, Activity
  recognition for ambient assisted living with videos, inertial units and
  ambient sensors, Sensors 21~(3) (2021) 768.

\bibitem{byabazaire2020using}
J.~Byabazaire, G.~O’Hare, D.~Delaney, Using trust as a measure to derive data
  quality in data shared iot deployments, in: 2020 29th International
  Conference on Computer Communications and Networks (ICCCN), IEEE, 2020, pp.
  1--9.

\bibitem{palmisani2021indoor}
J.~Palmisani, A.~Di~Gilio, M.~Viana, G.~de~Gennaro, A.~Ferro, Indoor air
  quality evaluation in oncology units at two european hospitals: Low-cost
  sensors for tvocs, pm2. 5 and co2 real-time monitoring, Building and
  Environment 205 (2021) 108237.

\bibitem{desouza2021distribution}
P.~deSouza, P.~L. Kinney, On the distribution of low-cost pm2. 5 sensors in the
  us: demographic and air quality associations, Journal of Exposure Science \&
  Environmental Epidemiology 31~(3) (2021) 514--524.

\bibitem{placidi2022capacitive}
P.~Placidi, N.~Papini, C.~V. Delle~Vergini, P.~Mezzanotte, A.~Scorzoni,
  Capacitive low-cost system for soil water content measurement in the iot
  precision agriculture, in: 2022 IEEE International Instrumentation and
  Measurement Technology Conference (I2MTC), IEEE, 2022, pp. 1--6.

\bibitem{theodorou2020network}
V.~Theodorou, M.-E. Xezonaki, Network slicing for multi-tenant edge processing
  over shared iot infrastructure, in: 2020 6th IEEE Conference on Network
  Softwarization (NetSoft), IEEE, 2020, pp. 8--14.

\bibitem{preuveneers2019big}
D.~Preuveneers, E.~Ilie-Zudor, Big data for context-aware applications and
  intelligent environments (2019).

\bibitem{LI2020111990}
D.~Li, Y.~Wang, J.~Wang, C.~Wang, Y.~Duan,
  \href{https://www.sciencedirect.com/science/article/pii/S0924424719308635}{Recent
  advances in sensor fault diagnosis: A review}, Sensors and Actuators A:
  Physical 309 (2020) 111990.
\newblock \href {http://dx.doi.org/https://doi.org/10.1016/j.sna.2020.111990}
  {\path{doi:https://doi.org/10.1016/j.sna.2020.111990}}.
\newline\urlprefix\url{https://www.sciencedirect.com/science/article/pii/S0924424719308635}

\bibitem{PALLEWATTA2022121}
S.~Pallewatta, V.~Kostakos, R.~Buyya,
  \href{https://www.sciencedirect.com/science/article/pii/S0167739X22000206}{Qos-aware
  placement of microservices-based iot applications in fog computing
  environments}, Future Generation Computer Systems 131 (2022) 121--136.
\newblock \href
  {http://dx.doi.org/https://doi.org/10.1016/j.future.2022.01.012}
  {\path{doi:https://doi.org/10.1016/j.future.2022.01.012}}.
\newline\urlprefix\url{https://www.sciencedirect.com/science/article/pii/S0167739X22000206}

\bibitem{dragoni2017microservices}
N.~Dragoni, S.~Giallorenzo, A.~L. Lafuente, M.~Mazzara, F.~Montesi,
  R.~Mustafin, L.~Safina, Microservices: yesterday, today, and tomorrow,
  Present and ulterior software engineering (2017) 195--216.

\bibitem{8816926}
K.~{Grueneberg}, B.~{Ko}, D.~{Wood}, X.~{Wang}, D.~{Steuer}, Y.~{Lim}, Iot data
  management system for rapid development of machine learning models, in: 2019
  IEEE International Conference on Cognitive Computing (ICCC), 2019, pp.
  59--63.
\newblock \href {http://dx.doi.org/10.1109/ICCC.2019.00021}
  {\path{doi:10.1109/ICCC.2019.00021}}.

\bibitem{xu2022amnis}
J.~Xu, B.~Palanisamy, Q.~Wang, H.~Ludwig, S.~Gopisetty, Amnis: Optimized stream
  processing for edge computing, Journal of Parallel and Distributed Computing
  160 (2022) 49--64.

\bibitem{8711204}
A.~Das, R.~Gupta, S.~Chakraborty, Motivating in-network fusion for smart
  infrastructure monitoring, in: 2019 11th International Conference on
  Communication Systems Networks (COMSNETS), 2019, pp. 513--516.
\newblock \href {http://dx.doi.org/10.1109/COMSNETS.2019.8711204}
  {\path{doi:10.1109/COMSNETS.2019.8711204}}.

\bibitem{shen2021temporal}
H.~Shen, W.~Hou, Y.~Zhu, S.~Zheng, S.~Ainiwaer, G.~Shen, Y.~Chen, H.~Cheng,
  J.~Hu, Y.~Wan, et~al., Temporal and spatial variation of pm2. 5 in indoor air
  monitored by low-cost sensors, Science of The Total Environment 770 (2021)
  145304.

\bibitem{giordano2021low}
M.~R. Giordano, C.~Malings, S.~N. Pandis, A.~A. Presto, V.~McNeill, D.~M.
  Westervelt, M.~Beekmann, R.~Subramanian, From low-cost sensors to
  high-quality data: A summary of challenges and best practices for effectively
  calibrating low-cost particulate matter mass sensors, Journal of Aerosol
  Science 158 (2021) 105833.

\bibitem{concas2021low}
F.~Concas, J.~Mineraud, E.~Lagerspetz, S.~Varjonen, X.~Liu, K.~Puolam{\"a}ki,
  P.~Nurmi, S.~Tarkoma, Low-cost outdoor air quality monitoring and sensor
  calibration: A survey and critical analysis, ACM Transactions on Sensor
  Networks (TOSN) 17~(2) (2021) 1--44.

\bibitem{petrich2020note}
C.~Petrich, I.~V. S{\ae}ther, N.~P. Dang, {\O}.~Kleven, M.~O'Sadnick, A note on
  remote temperature measurements with ds18b20 digital sensors, in: PROCEEDINGS
  OF THE 25th INTERNATIONAL SYMPOSIUM ON ICE, 2020.

\bibitem{de2016benchmark}
B.~de~Bruijn, T.~A. Nguyen, D.~Bucur, K.~Tei, Benchmark datasets for fault
  detection and classification in sensor data., in: SENSORNETS, 2016, pp.
  185--195.

\bibitem{jan2017sensor}
S.~U. Jan, Y.-D. Lee, J.~Shin, I.~Koo, Sensor fault classification based on
  support vector machine and statistical time-domain features, IEEE Access 5
  (2017) 8682--8690.

\bibitem{ping2007voting}
W.~Ping, A voting strategy for n-version program based on fuzzy clustering, in:
  2007 8th International Conference on Electronic Measurement and Instruments,
  IEEE, 2007, pp. 2--666.

\bibitem{kuo2003optimal}
W.~Kuo, M.~J. Zuo, Optimal reliability modeling: principles and applications,
  John Wiley \& Sons, 2003.

\bibitem{gu2017new}
X.~Gu, P.~P. Angelov, D.~Kangin, J.~C. Principe, A new type of distance metric
  and its use for clustering, Evolving Systems 8~(3) (2017) 167--177.

\bibitem{choudhury2019proactive}
B.~Choudhury, S.~Choudhury, A.~Dutta, A proactive context-aware service
  replication scheme for adhoc iot scenarios, IEEE Transactions on Network and
  Service Management 16~(4) (2019) 1797--1811.

\bibitem{DASILVA20191041}
M.~P. {da Silva}, A.~L. Gonçalves, M.~A.~R. Dantas,
  \href{https://www.sciencedirect.com/science/article/pii/S0167739X18314079}{A
  conceptual model for quality of experience management to provide
  context-aware ehealth services}, Future Generation Computer Systems 101
  (2019) 1041--1061.
\newblock \href
  {http://dx.doi.org/https://doi.org/10.1016/j.future.2019.07.033}
  {\path{doi:https://doi.org/10.1016/j.future.2019.07.033}}.
\newline\urlprefix\url{https://www.sciencedirect.com/science/article/pii/S0167739X18314079}

\bibitem{gnauck2004interpolation}
A.~Gnauck, Interpolation and approximation of water quality time series and
  process identification, Analytical and bioanalytical chemistry 380~(3) (2004)
  484--492.

\bibitem{lepot2017interpolation}
M.~Lepot, J.-B. Aubin, F.~H. Clemens, Interpolation in time series: An
  introductive overview of existing methods, their performance criteria and
  uncertainty assessment, Water 9~(10) (2017) 796.

\bibitem{9318810}
M.~Fikri, S.~Herdjunanto, A.~Cahyadi, On the performance similarity between
  exponential moving average and discrete linear kalman filter, in: 2019 Asia
  Pacific Conference on Research in Industrial and Systems Engineering
  (APCoRISE), 2019, pp. 1--5.
\newblock \href {http://dx.doi.org/10.1109/APCoRISE46197.2019.9318810}
  {\path{doi:10.1109/APCoRISE46197.2019.9318810}}.

\bibitem{QIU2022241}
S.~Qiu, H.~Zhao, N.~Jiang, Z.~Wang, L.~Liu, Y.~An, H.~Zhao, X.~Miao, R.~Liu,
  G.~Fortino,
  \href{https://www.sciencedirect.com/science/article/pii/S1566253521002311}{Multi-sensor
  information fusion based on machine learning for real applications in human
  activity recognition: State-of-the-art and research challenges}, Information
  Fusion 80 (2022) 241--265.
\newblock \href
  {http://dx.doi.org/https://doi.org/10.1016/j.inffus.2021.11.006}
  {\path{doi:https://doi.org/10.1016/j.inffus.2021.11.006}}.
\newline\urlprefix\url{https://www.sciencedirect.com/science/article/pii/S1566253521002311}

\bibitem{candanedo2016accurate}
L.~M. Candanedo, V.~Feldheim, Accurate occupancy detection of an office room
  from light, temperature, humidity and co2 measurements using statistical
  learning models, Energy and Buildings 112 (2016) 28--39.

\bibitem{benammar2018modular}
M.~Benammar, A.~Abdaoui, S.~H. Ahmad, F.~Touati, A.~Kadri, A modular iot
  platform for real-time indoor air quality monitoring, Sensors 18~(2) (2018)
  581.

\bibitem{nunes2018elimination}
L.~H. Nunes, J.~C. Estrella, C.~Perera, S.~Reiff-Marganiec, A.~C. Delbem, The
  elimination-selection based algorithm for efficient resource discovery in
  internet of things environments, in: 2018 15th IEEE annual consumer
  communications \& networking conference (CCNC), IEEE, 2018, pp. 1--7.

\bibitem{XIAO201923}
F.~Xiao,
  \href{https://www.sciencedirect.com/science/article/pii/S1566253517305584}{Multi-sensor
  data fusion based on the belief divergence measure of evidences and the
  belief entropy}, Information Fusion 46 (2019) 23--32.
\newblock \href
  {http://dx.doi.org/https://doi.org/10.1016/j.inffus.2018.04.003}
  {\path{doi:https://doi.org/10.1016/j.inffus.2018.04.003}}.
\newline\urlprefix\url{https://www.sciencedirect.com/science/article/pii/S1566253517305584}

\bibitem{liggins2017handbook}
M.~Liggins~II, D.~Hall, J.~Llinas, Handbook of multisensor data fusion: theory
  and practice, CRC press, 2017.

\bibitem{ramirez2017survey}
S.~Ram{\'\i}rez-Gallego, B.~Krawczyk, S.~Garc{\'\i}a, M.~Wo{\'z}niak,
  F.~Herrera, A survey on data preprocessing for data stream mining: Current
  status and future directions, Neurocomputing 239 (2017) 39--57.

\bibitem{kenda2019streaming}
K.~Kenda, B.~Ka{\v{z}}i{\v{c}}, E.~Novak, D.~Mladeni{\'c}, Streaming data
  fusion for the internet of things, Sensors 19~(8) (2019) 1955.

\bibitem{yin2020novel}
Y.~Yin, B.~Xu, H.~Cai, H.~Yu, A novel temporal and spatial panorama stream
  processing engine on iot applications, Journal of Industrial Information
  Integration 18 (2020) 100143.

\bibitem{peros2021ermis}
S.~Peros, W.~Joosen, D.~Hughes, Ermis: a middleware for bridging data
  collection and data processing in iot streaming applications, in: 2021 17th
  International Conference on Distributed Computing in Sensor Systems (DCOSS),
  IEEE, 2021, pp. 259--266.

\bibitem{garcia2016big}
S.~Garc{\'\i}a, S.~Ram{\'\i}rez-Gallego, J.~Luengo, J.~M. Ben{\'\i}tez,
  F.~Herrera, Big data preprocessing: methods and prospects, Big Data Analytics
  1~(1) (2016) 1--22.

\bibitem{costa2021goat}
F.~S. Costa, S.~M. Nassar, M.~A. Dantas, Goat: A sensor ranking approach for
  iot environments., in: CLOSER, 2021, pp. 169--177.

\bibitem{saito2021determinant}
Y.~Saito, T.~Nonomura, K.~Yamada, K.~Nakai, T.~Nagata, K.~Asai, Y.~Sasaki,
  D.~Tsubakino, Determinant-based fast greedy sensor selection algorithm, IEEE
  Access 9 (2021) 68535--68551.

\bibitem{bharti2019value}
S.~Bharti, K.~K. Pattanaik, P.~Bellavista, Value of information based sensor
  ranking for efficient sensor service allocation in service oriented wireless
  sensor networks, IEEE Transactions on Emerging Topics in Computing.

\bibitem{bharti2020optimal}
M.~Bharti, R.~Kumar, S.~Saxena, H.~Jindal, Optimal resource selection framework
  for internet-of-things, Computers \& Electrical Engineering 86 (2020) 106693.

\bibitem{babu2019context}
K.~R. Babu, M.~V. Prathap, P.~Samuel, Context aware reliable sensor selection
  in iot, International Journal of Intelligent Systems Technologies and
  Applications 18~(1-2) (2019) 34--51.

\bibitem{perera2013context}
C.~Perera, A.~Zaslavsky, P.~Christen, M.~Compton, D.~Georgakopoulos,
  Context-aware sensor search, selection and ranking model for internet of
  things middleware, in: 2013 IEEE 14th international conference on mobile data
  management, Vol.~1, IEEE, 2013, pp. 314--322.

\bibitem{kuhlmann2021smartphone}
T.~Kuhlmann, P.~Garaizar, U.-D. Reips, Smartphone sensor accuracy varies from
  device to device in mobile research: The case of spatial orientation,
  Behavior research methods 53 (2021) 22--33.

\bibitem{tsarev2019fuzzy}
R.~Tsarev, M.~Durmu{\c{s}}, I.~{\"U}stoglu, V.~Morozov, A.~Pupkov, Fuzzy voting
  algorithms for n-version software, in: Journal of Physics: Conference Series,
  Vol. 1333, IOP Publishing, 2019, p. 032087.

\bibitem{tsarev2018classification}
R.~Y. Tsarev, M.~S. Durmu{\c{s}}, I.~{\"U}stoglu, V.~Morozov, Classification of
  voting algorithms for n-version software, in: Journal of Physics: Conference
  Series, Vol. 1015, IOP Publishing, 2018, p. 042060.

\bibitem{yan2020collaborative}
J.~Yan, W.~Pu, S.~Zhou, H.~Liu, Z.~Bao, Collaborative detection and power
  allocation framework for target tracking in multiple radar system,
  Information Fusion 55 (2020) 173--183.

\bibitem{ozcan2020accurate}
M.~{\"O}zcan, F.~Aliew, H.~G{\"o}rg{\"u}n, Accurate and precise distance
  estimation for noisy ir sensor readings contaminated by outliers, Measurement
  156 (2020) 107633.

\bibitem{wen2014accuracy}
H.~Wen, Z.~Xiao, A.~Markham, N.~Trigoni, Accuracy estimation for sensor
  systems, IEEE Transactions on Mobile Computing 14~(7) (2014) 1330--1343.

\bibitem{mukhopadhyay2008model}
S.~Mukhopadhyay, C.~Schurgers, D.~Panigrahi, S.~Dey, Model-based techniques for
  data reliability in wireless sensor networks, IEEE Transactions on Mobile
  Computing 8~(4) (2008) 528--543.

\bibitem{ni2009sensor}
K.~Ni, N.~Ramanathan, M.~N.~H. Chehade, L.~Balzano, S.~Nair, S.~Zahedi,
  E.~Kohler, G.~Pottie, M.~Hansen, M.~Srivastava, Sensor network data fault
  types, ACM Transactions on Sensor Networks (TOSN) 5~(3) (2009) 1--29.

\bibitem{noshad2019fault}
Z.~Noshad, N.~Javaid, T.~Saba, Z.~Wadud, M.~Q. Saleem, M.~E. Alzahrani, O.~E.
  Sheta, Fault detection in wireless sensor networks through the random forest
  classifier, Sensors 19~(7) (2019) 1568.

\bibitem{zidi2017fault}
S.~Zidi, T.~Moulahi, B.~Alaya, Fault detection in wireless sensor networks
  through svm classifier, IEEE Sensors Journal 18~(1) (2017) 340--347.

\bibitem{ghosh2020fault}
N.~Ghosh, R.~Paul, S.~Maity, K.~Maity, S.~Saha, Fault matters: Sensor data
  fusion for detection of faults using dempster--shafer theory of evidence in
  iot-based applications, Expert Systems with Applications 162 (2020) 113887.

\bibitem{jiwei2018reliability}
Q.~Jiwei, Z.~Jianguo, M.~Yupeng, Reliability analysis based on the principle of
  maximum entropy and dempster--shafer evidence theory, Journal of Mechanical
  Science and Technology 32~(2) (2018) 605--613.

\end{thebibliography}

\end{document}